\documentclass[acmtog]{acmart}
\acmSubmissionID{434} 
\usepackage{booktabs} 
\usepackage{cuted}
\usepackage{xcolor, colortbl}
\usepackage{wrapfig}

\usepackage[ruled]{algorithm2e} 

\SetAlFnt{\small}
\SetAlCapFnt{\small}
\SetAlCapNameFnt{\small}
\SetAlCapHSkip{0pt}

\acmJournal{TOG}

\usepackage{hyperref}
\usepackage{url}

\usepackage{multirow}
\usepackage{overpic}
\usepackage[T1]{fontenc}

\newcommand{\eg}{\emph{e.g.}\xspace}

\graphicspath{{figs/}}
\begin{document}

\title{A Variational Loop Shrinking Analogy for Handle and Tunnel
Detection and Reeb Graph Construction on Surfaces  }

\author{Alexander Weinrauch}
\email{alexander.weinrauch@icg.tugraz.at}
\affiliation{%
  \institution{Graz University of Technology}
  \country{Austria}
  }
\affiliation{%
 \institution{Max Planck Institute for Informatics}
 \country{Germany}  
}

\author{Hans-Peter Seidel}
\email{hpseidel@mpi-inf.mpg.de}
\affiliation{%
 \institution{Max Planck Institute for Informatics, Saarland Informatics Campus}
 \country{Germany}
}

\author{Daniel Mlakar}
\email{daniel.mlakar@icg.tugraz.at}
\affiliation{%
  \institution{Graz University of Technology}
  \country{Austria}
}

\author{Markus Steinberger}
\email{steinberger@icg.tugraz.at}
\affiliation{%
  \institution{Graz University of Technology}
  \country{Austria}
}

\author{Rhaleb Zayer}
\email{rhaleb.zayer@mpi-inf.mpg.de}
\affiliation{%
 \institution{Max Planck Institute for Informatics, Saarland Informatics Campus}
 \city{Saarbrücken}
 \country{Germany}
}


\begin{abstract}

The humble \emph{loop shrinking property} played a central role in the inception of modern topology but it has been eclipsed by more abstract algebraic formalism. This is particularly true in the context of detecting relevant non-contractible loops on surfaces where elaborate homological and/or graph theoretical constructs are favored in algorithmic solutions.
In this work, we devise a variational analogy to the loop shrinking property and show that it yields a simple, intuitive, yet powerful solution allowing a streamlined treatment of the problem of handle and tunnel loop detection. Our formalization tracks the evolution of a diffusion front randomly initiated on a single location on the surface. Capitalizing on a diffuse interface representation combined with a set of rules for concurrent front interactions, we develop a dynamic data structure for tracking the evolution on the surface encoded as a sparse matrix which serves for performing both diffusion numerics and loop detection and acts as the workhorse of our fully parallel implementation. The substantiated results suggest our approach outperforms state of the art and robustly copes with highly detailed geometric models. As a byproduct, our approach can be used to construct \emph{Reeb graphs} by diffusion thus avoiding commonly encountered issues when using Morse functions.
\end{abstract}

\ccsdesc[500]{Computing methodologies~Shape analysis}
\ccsdesc[500]{Computing methodologies~Massively parallel algorithms}

\keywords{Handle loop, Tunnel loop, Reeb graph }


\begin{teaserfigure}
	\centering
	\includegraphics[width=\textwidth]{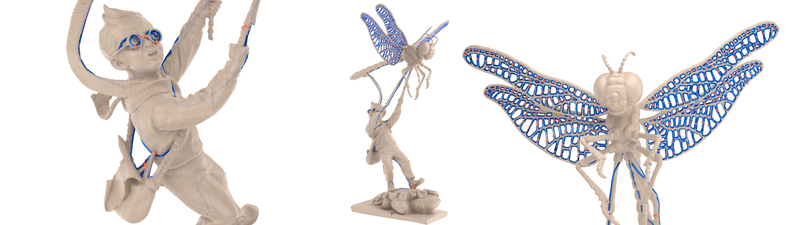}
	\caption{Computed handle and tunnel loops on the Dragon Tamer model.}
	\label{fig:teaser}
\end{teaserfigure}

\maketitle

\section{Introduction}
\label{sec:intro}
Probably one of the most inspiring achievements of topology is the classification theorem for surfaces, which establishes equivalence classes based on the \emph{Euler characteristic} and \emph{orientability}.  This development triggered  a formidable effort focalized on the so called  \emph{Poincar\'e conjuncture} aiming at extending classification to higher dimensions and in particular to the $3-$manifold setting and captured the imagination of generations of mathematicians and the general public alike.
Loosely speaking the classification theorem for surfaces (2-manifolds) tells us that any orientable surface is equivalent to a sphere with a certain number of "handles" sewn onto it. In this respect, the surface can be constructed from a sphere by \emph{topological surgery}, which can be understood as a set of cutting, stitching and deforming operations.

Although a sound theoretical foundation of the subject matter has been laid out in algebraic topology, see e.g.~\cite{munkres1984elements}, localization of geometrically meaningful handles on surfaces remains a highly challenging topic. Over the last decades, several algorithms for detecting non trivial loops on general graphs have been proposed in computational geometry, see e.g. \cite{erickson2012combinatorial} and the references therein, but most of these results remain theoretical and have not turned into efficient implementations. The topic of extracting such topological features is not only relevant as a theoretical pursuit but has fundamental applications in geometry processing, covering  tasks such as mesh parametrization, mesh repair, and feature recognition, and spills beyond to fields such as biotechnology and bioinformatics, see e.g.~\cite{brezovsky2013software,voss20103v},  therefore there is need for efficient practical solutions.
The pioneering efforts of Dey and coworkers~\shortcite{dey2007,dey2013} formalized the notion of geometrically meaningful topological features such as handle and tunnel loops and presented practical solutions using intermediate structures relying on \emph{persistence homology} and \emph{Reeb graphs}. Nonetheless, the cost of all intermediate constructs and their limitations impedes performance and intensifies memory usage.

Our goal is to provide  a simple, intuitive, yet efficient strategy for detection of handle and tunnel loops. To this end, we re-examine the problem in the light of the humble loop shrinking property which lies behind Poincar\'e's intuition and we seek to develop a variational analog to it which would allow us to evolve freely on the surface. In doing so, we avoid the difficulties encountered by existing methods in their efforts to marry homotopy classes, elaborate graph constructs, and practical geometric requirements.

\begin{figure}[h]
	\centering
	\includegraphics[width=\linewidth]{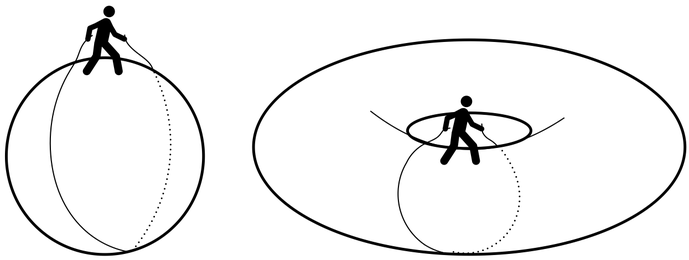}
	\caption{Loop shrinking property illustrated on simple geometric objects.}
	\label{fig:walker}
\end{figure}

Consider a person walking on a given surface while holding a sufficiently long thread from both ends, as illustrated in Figure~\ref{fig:walker}. Had the person been on a sphere, she would be able to spool it back. On the other hand, if the person is on a torus, it won't  be able to re-spool the thread because it passes through the 2-dimensional hole of the torus.

\begin{figure}[h]
	\centering
	\includegraphics[width=.48\linewidth]{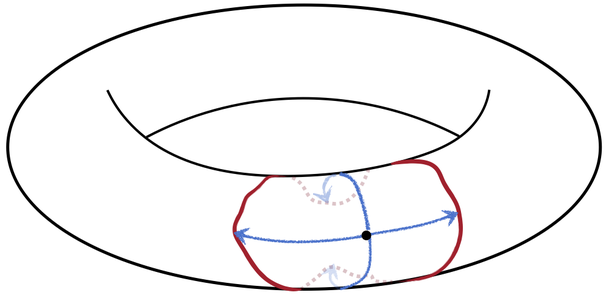}
	\includegraphics[width=.48\linewidth]{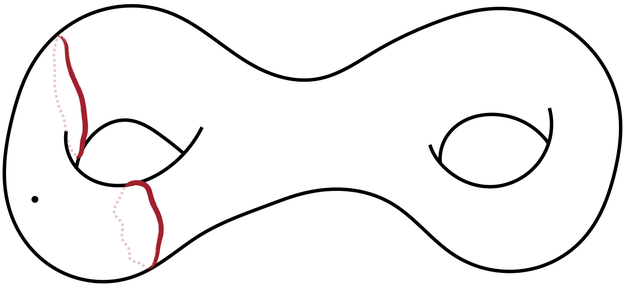}
	\includegraphics[width=.48\linewidth]{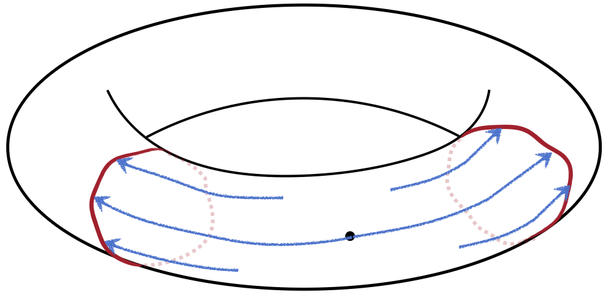}
	\includegraphics[width=.48\linewidth]{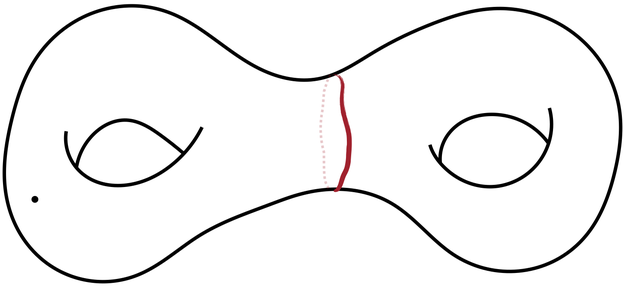}
	\includegraphics[width=.48\linewidth]{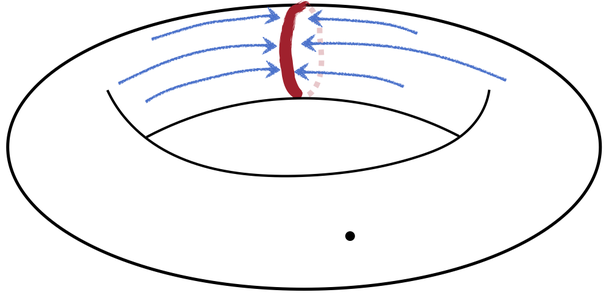}
	\includegraphics[width=.48\linewidth]{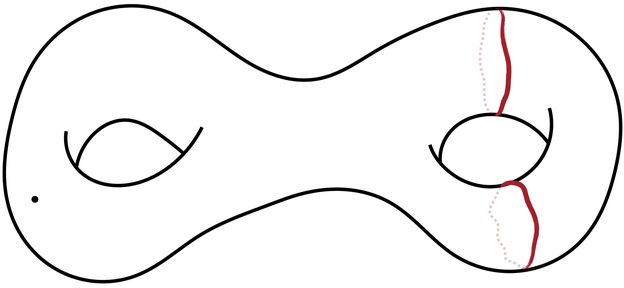}

	\caption{Handle loop discovery through splitting and collision of diffusion fronts on a simple torus (left) and a double torus (right). }
	\label{fig:illustration_loops}
\end{figure}

\begin{figure}[t]
	\centering
	\includegraphics[width=0.32\linewidth]{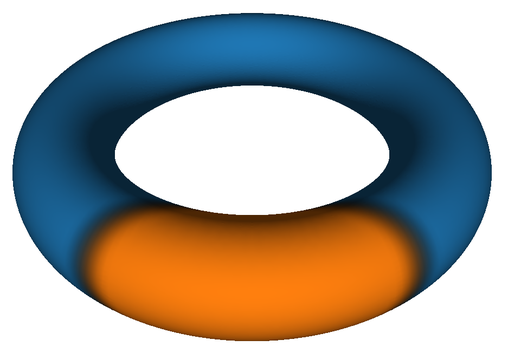}
	\includegraphics[width=0.32\linewidth]{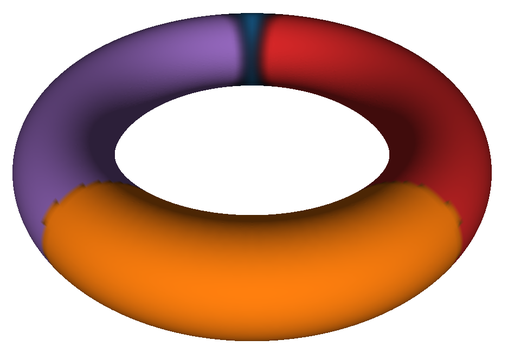}
	\includegraphics[width=0.32\linewidth]{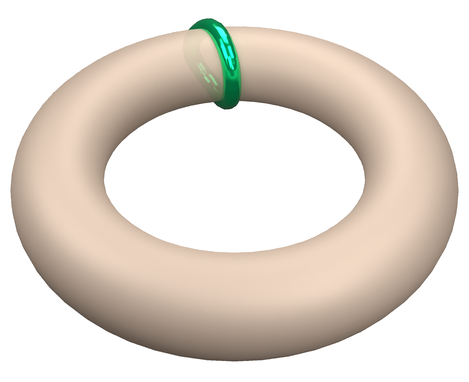}
	\includegraphics[width=0.32\linewidth]{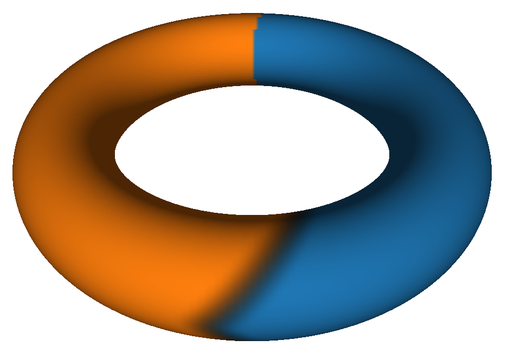}
	\includegraphics[width=0.32\linewidth]{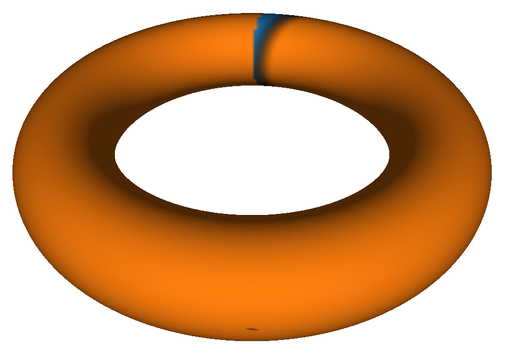}
	\includegraphics[width=0.32\linewidth]{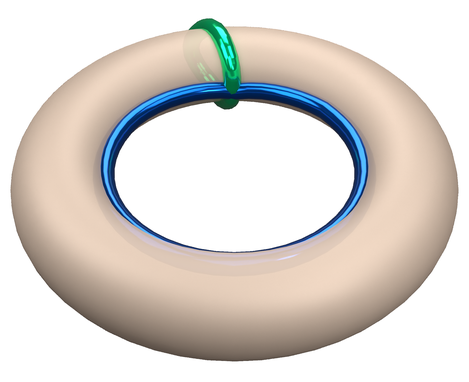}
	\includegraphics[width=0.32\linewidth]{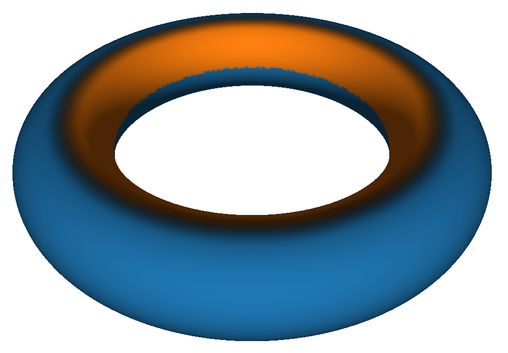}
	\includegraphics[width=0.32\linewidth]{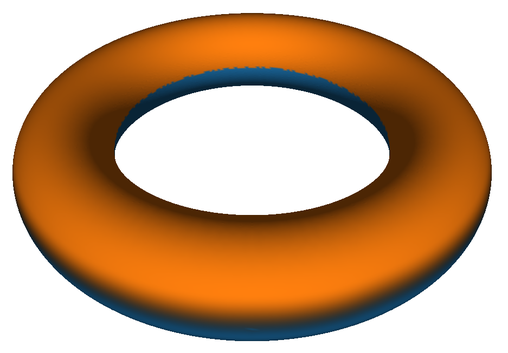}
	\includegraphics[width=0.32\linewidth]{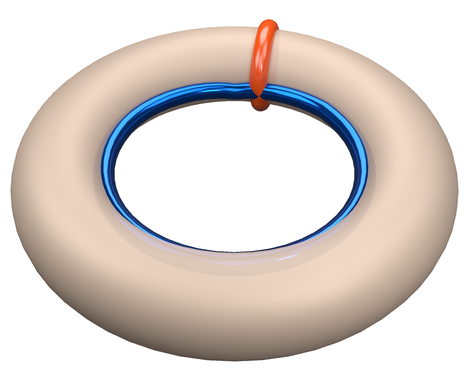}
	\caption{In the initial pass (top), the field interface (rim of the orange region) splits into two different components (top-left), both of which propagate till they reach each other (top-middle) yielding an initial handle estimate (top-right). In the second pass (middle), a diffusion initiated leftwards of the handle (middle-left) progresses till it reaches the opposite side of the handle (middle-middle). A streamline tracing against the diffusion gradient field yields the tunnel loop (middle-right). Similarly, in the third pass (bottom), a diffusion from the tunnel loop upwards (bottom-left) reaches the opposite side of the tunnel (bottom-middle) yielding a refined handle loop (bottom-right).}
	\label{fig:torus_initial_pass}
\end{figure}

To capture the essence of the loop shrinking property, we consider a diffusion process starting from an arbitrary location $p$ on the surface. If we are on the surface of a sphere, the advancing front will grow steadily but eventually it will start shrinking and vanish to one point similar to the thread losing hold. On the other hand, if we are on the surface of a torus, the growing patch will initially have a single boundary (Figure~\ref{fig:illustration_loops}-left-top), and eventually, it will meet itself as it folds like a cannoli yielding a tunnel like region with two separate boundaries (Figure~\ref{fig:illustration_loops}-left-middle). At this point, we can only confirm that we are evolving on a tubular structure but we cannot infer the existence of a handle. Only at the moment when the two advancing fronts meet (Figure~\ref{fig:illustration_loops}-left-bottom), then a handle loop is detected. A tunnel loop can be obtained as a streamline by tracing back from the handle loop along the diffusion gradient,  see Figure \ref{fig:torus_initial_pass}. Clearly here, the number of possible tunnel loops is large and all of them will pass through the initial point $p$, so there is is no reason to expect them to exhibit some geometric optimality. This can be remedied by performing a diffusion initiated from the ring of the handle loop (Figure~\ref{fig:illustration_loops}-left-bottom). Once the diffusion converges we can trace multiple streamlines sampled along the handle loop, thus obtaining a set of tunnel loops and we can then select the shortest.

This idea extends naturally to higher genus settings. For instance, let us consider the case of the double torus in Figure~\ref{fig:illustration_loops}-right-top, the diffusion from the point marked on the surface will behave initially in a similar manner to the torus case, but as the two fronts merge (Figure~\ref{fig:illustration_loops}-right-middle), the resulting single front will eventually split into two fronts (Figure~\ref{fig:illustration_loops}-right-bottom) which would meet each other again yielding the second handle loop.

Despite the inherent simplicity of the process described above, special attention needs to be paid to front tracking, bookkeeping, memory usage and performance in order to build a practical solution suitable for large data sets and complex surface layouts. A possible way forward is to use the level set method for carrying out the front tracking of the diffusion on the surface, however the sharp interface would require substantial bookkeeping and costly computations and checks on top of commonly known issues of the method pertaining to the distance function evaluation and re-initialization.
Instead, we rely on the recent layered fields approach proposed as an alternative model to the ubiquitous Voronoi diagrams for modeling natural tessellation modeling~\cite{zayer2018}. We use a narrow band to represent the advancing front to avoid discontinuities across the front.  We regard the propagation as field evolving on a layer on the surface. The splitting of the interface corresponds to the splitting of the field into two different fields evolving on different layers. For this purpose, we develop a dynamic approach to layer creation and layer collision which allows us to keep track of all the handles and tunnels on non trivial higher genus surfaces.

We represent the different layers as rows of a sparse matrix which both allows for carrying numerical diffusion computations and the tracking of topological features. As our geometry processing operations are channeled through linear algebra we take full advantage of parallel linear algebra primitives. In particular, our implementation operates fully on graphics hardware. A windfall of our approach is the ability to generate a Reeb graph simply by performing diffusion initiated at the detected handle loops. In this way, the commonly encountered shortcomings of working with height based morse functions are avoided in the first place.

In summary, this work makes the following contributions:
\begin{itemize}
\item Novel variational abstraction for the loop shrinking property
\item Succinct diffuse interface Model for identifying handle and tunnel loops and capturing front propagation and collision  
\item	Dynamic sparse matrix representation for bookkeeping and processing
\item Fully parallel algorithmic realization on graphics hardware
\item Simple and intuitive Reeb graph extraction
\end{itemize}

\paragraph {Previous work}
The body of work on extracting topological primitives on surfaces spans efforts across computational geometry and mesh processing. Theoretical algorithms for extracting non-trivial loops on surfaces have been proposed, \eg, \cite{Erickson2005,kutz2006,deVerdiere2006}. However there are no existing realizations of these theoretical efforts and no guarantees that the resulting loops are geometrically meaningful handles or tunnels. In the context of mesh simplification, several heuristics for identifying small handles have been proposed, \eg,~\cite{elsana1997,Guskov2001}.
The use of intermediate graphs for non-trivial loop identification has been explored in terms of \emph{Reeb graphs} in \cite{Shattuck2001,wood2004} and in terms of medial axis in \cite{Zhou2007}.

More closely related to our work are the efforts directly targeting the localization of handle and tunnel loops. Dey and coworkers addressed the problem by relying on persistence homology in~\cite{Dey2008}, and then later on by making use of Reeb graphs~\cite{dey2013}. In the former, the intermediate tetrahedral tessellation required for carrying out homology computations poses several challenges both to feasibility and performance and bloats the problem size unnecessarily. Dropping persistence homology, in the latter, in favor of Reeb graphs improved both performance and scope, but computational cost and memory requirements remain considerable.

A data structure based on tree-cotree decomposition which allows for constructing generators of fundamental groups of surfaces was proposed by Eppstein~\shortcite{Eppstein2003}. Adjusting edge weights of the tree-cotree decomposition based on principal curvature directions, \cite{Diaz-Gutierrez2009} attempt to inject geometric meaning into those graph cycles by to steering them to align with those directions. These methods are prone to produce redundant non-trivial cycles which require post processing. An iterative approach alternating between principal directions to find ``good'' cycles was proposed in~\cite{Chen2018}. It should be noted that curvature directions on general surfaces do not necessarily match the targeted cycle directions as the surface can be heavily decorated, see for instance, the dragons models in Figure~\ref{fig:results_set} and this limits the scope to overly smooth meshes.

Without loss of generality, the above discussed methods do not operate directly on the $2$D manifold but require one or multiple intermediate structures. Subsequently, performance and results quality depend on the quality of these underlaying structures. For instance, in \cite{dey2013}, the Reeb graph quality depends among other thing on the choice of the initial direction as usually the associated morse function is some height function. Likewise,~\cite{Diaz-Gutierrez2009,Chen2018} depend on principal directions estimations which are generally local, sensitive to surface fluctuations and noise. More importantly, the resulting algorithmic pipeline is not streamlined and therefore pose further challenges to code vectorization in view of the ubiquitous parallelism in modern hardware infrastructure.

\section{Mathematical representation and data structure}
\label{sec:math}

\paragraph{General setting.} A majority of existing methods operate in the combinatorial manifold setting where the computation of shortest homology generators is relatively simpler than the more geometrically meaningful piecewise linear manifold. While this simplification offers certain algorithmic guarantees and computational advantages, the results depends largely on the quality of the underlying mesh edges. The challenges of working directly on the piecewise linear manifold have long been recognized, for instance,~\cite{Erickson2005} discuss using level-sets for speeding up shortest path computations but acknowledge that “even the simpler problem of finding the shortest non-separating cycle in a piecewise-linear manifold appears to be open”. 

Similar to~\cite{dey2013}, our work seeks to find geometrically meaningful generating pairs, meaningful is to be understood in the sense of shortest loops. In our illustrations, e.g., Figure~\ref{fig:illustration_loops}, as in most common objects, the handle loop is way shorter than the tunnel loop (think a mug handle for instance), so for the sake of presentation flow, we presume in the ensuing discussion that a handle loops are found before their tunnel counterparts. Of course, it is possible to imagine scenarios where the opposite happens, e.g., deep screw holes in the drill in Figure~\ref{fig:result_napoleon}) but this does not affect the pair extraction per se and the distinction between handle and tunnel loops will be discussed aside.

\paragraph{Diffuse interface.} 
While our approach is driven by diffusion, a pure diffusion alone is not sufficient for tracking different advancing fronts and resolving front collisions. In fact, it would lead simply to a smeared interface which would eventually end up as a flat curve (merged fronts).  This issue gets further amplified on higher genus surfaces as multiple branchings yield multiple concurrent fronts.
In order to resolve these issues and to avoid the numerical problems commonly encountered with sharp interface such as derivative discontinuities, we adopt a diffuse interface approach where the advancing front is not a sharp line as in the level set method but a narrow band. 
\setlength\intextsep{0pt}
\setlength{\columnsep}{3pt}%
\begin{wrapfigure}{r}{0.36\linewidth}
  \centering
    \includegraphics[width=\linewidth]{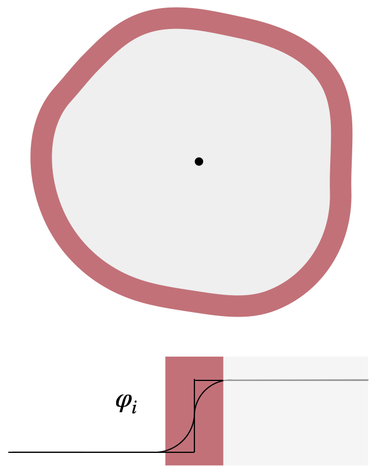}
\end{wrapfigure}
\setlength\intextsep{8pt}
\setlength{\columnsep}{16pt}%
In this way the growing patch can be defined by a function $\varphi_i$ define dover the whole mesh and  which takes value $1$ inside the patch and $0$ outside the cell and is smoothly varying on the narrow  a narrow band where $0 < \varphi_i <1$. In this way, it is easy to localize the narrow band by simple inspection of field values. Furthermore, at any time, it is possible to extract a sharp boundary as a level set of the narrow band.

\paragraph{Mathematical characterization.} In our context, front splitting due to branching, see e.g. Figure~\ref{fig:illustration_loops}, leads to the creation of new separate regions and front collision requires colliding fronts to freeze thus blocking each other. These behavior patterns are similar to the natural tessellation model proposed in~\cite{zayer2018} which describes the evolution of the interface of multiple cells growing simultaneously on a surface and yields Voronoi-like surface tessellations.

Within this setting, we can regard the diffusion of $n$ individual regions or cells as fields evolving on different layers and their interactions are governed by a set of requirements and weighted energy terms which drive the overall cell interactions:
\begin{enumerate}
\item Partition of unity (the different fields $\varphi_i$ sum up to $1$ at any given location on the surface). 
\item Integrity and locality of the individual cells in terms of the weighted product $w_{ij}\varphi_i\ \varphi_j$. The interaction does not cause a cell to disintegrate or break.
\item Distinguishability of the individual narrow bands in terms of the weighted gradients $-\frac{1}{2}a_{ij}\nabla\varphi_i\ \nabla\varphi_j$. Narrow bands corresponding to each cell do not merge and flatten out.
\item Behavior of narrow bands under contact which we model by a certain function $g$ which will be discussed shortly.
\end{enumerate}

We account for the partition of unity constraint by introducing the Lagrange multiplier $\lambda$. We define the global energy of the model on the surface $S$ as a combination of three remaining terms discussed above. The $\varphi_i$'s can be treated as independent variables, and the arising Lagrangian then reads
\begin{equation}\label{equ:Lagrange}
    \Gamma=\int_S \sum_{i=1}^n \sum_{j=i+1}^n (w_{ij} \varphi_i\varphi_j -\frac{a_{ij}}{2}\nabla\varphi_i \nabla \varphi_j + g) +\lambda (\sum_{i=1}^n \varphi_i -1) \quad ds,
\end{equation}
where the integral spans the surface $S$ and the summation is over the number of cells or regions $n$ .

Following~\cite{zayer2018}, the minimization of the Lagrangian associated with the energy arising from the above considerations yields the following time dependent equation which describes the evolution of the individual fields 
\begin{multline}\label{equ:phitimeDerivativeMain}
\dot{\varphi_i}=-\sum_{j=1}^n \frac{\mu}{n}\biggl(\sum_{k=1}^n\left[(w_{ik}-w_{jk})\varphi_k + \frac{1}{2}(a_{ik}^2-a_{jk}^2)\nabla^2 \varphi_k) \right] \\+(\frac{\partial g}{\partial \varphi_i}-\frac{\partial g}{\partial \varphi_j})\biggr).
\end{multline}
As observed in the above equation, We do not need to define $g$ explicitly, we only need to define the difference term $(\frac{\partial g}{\partial \varphi_i}-\frac{\partial g}{\partial \varphi_j})$. The function $g$ encodes the desired behavior when two boundary bands meet each other. In the simple case, where only two bands ${\varphi_i, \varphi_j}$ meet, we have by virtue  of partition of unity, $\varphi_j=1-\varphi_i$. In practice , we seek a function that produces a symmetric behavior in  the interval $(0,1)$ and vanishes at $0$ and $1$ and does not change sign. One such function is $\sqrt{\varphi_i (1-\varphi_i)}$, in this way, we have 
\begin{equation*}\label{equ:phitimeDerivativeMain}
\begin{tiny}
\dot{\varphi_i}=-\sum_{j=1}^n \frac{\mu_{ij}}{n}\biggl(\sum_{k=1}^n\left[(w_{ik}-w_{jk})\varphi_k + \frac{1}{2}(a_{ik}^2-a_{jk}^2)\nabla^2 \varphi_k \right] -e_{ij}\sqrt{\varphi_i\varphi_j} \biggr).
\end{tiny}
\end{equation*}

In the above equation, we can recognize the pure diffusion term characterized by the Laplacian which guarantees the smoothness of the growing interface. The first term in the equation, ensures a stable and stationary interface and therefore balances the effect of the diffusion term. The last term characterizes the driving force of the interface between the $i$th and the $j$th cell.

The growth rate is controlled by the mobility term $\mu_{ij}$. The solution can then be carried by a simple explicit Euler stepping scheme
$\varphi_i(t+\Delta t)=\varphi_i(t)+\dot{\varphi_i}\Delta t$. In all our experiments we used a constant time step. For a discussion of the effect of the different time step the reader is referred to ~\cite{zayer2018} or standard numerical methods textbooks.

\paragraph{Implementation details.} In our discretization on surface meshes, we uses the standard linear finite element approximation. The different fields can be thought of as individual layers on the surface and can be numerically encoded as a sparse matrix $\Phi$ where each column represents a vertex of the original input mesh and each row represents a field $\varphi_i$, i.e. a region. This representation is convenient in the sense that we can precompute the Laplacian matrix of the mesh and at each time step the evaluation of the Laplacian of the fields amount to a sparse matrix matrix multiplication with $\Phi$. The Laplacian of $\varphi_i$  is just the $i$th row in the resulting product. This formulation offers a clear advantage in view of efficient parallel implementation. Furthermore, checking if two cells share a boundary amounts their corresponding rows in $\Phi$ and looking for common nonzeros along the columns (vertices). To steer the field evolution, an additional base layer is initialized to $1$ at all vertices. When a region is initialized, we set the corresponding vertices inside to $1$ and to zero outside.
To help enforce partition of unity, especially across borders, the field $\Phi$ is normalized by dividing each column by the sum of its components. This normalization is further enforced after each time step. Furthermore, instead of a costly explicit tracking of cross-cell interactions at each iteration, we use the base layer for carrying this task implicitly.
When two boundary bands move towards each other (or multiple bands at a junction), the base layer between them erodes (the corresponding vertices in $\Phi$ are no longer $1$). With this in mind, both $e_{ij}$ and the mobility term $\mu_{ij}$ are set to zero except when either $i$ or $j$ index the base layer. We used $\frac{1}{30}$ and  $\frac{1}{4}$ resp. in our current implementation. The values of the gradient energy coefficients  $a_{i,j}$ are set to $\frac{1}{25}$, and the penalty term $w_{i,j}$ to $\frac{a_{i,j}}{5}$ for $i\neq j$, and to $0$ otherwise.

\section{Variational Loop Shrinking}
\label{sec:algo_details}

From the outline given in the introduction, our method can be regarded as sequence of three distinct passes. The initial pass registers the number of handle and tunnel pairs and gives an initial estimate for each handle loop. In the second pass, tunnel loops corresponding to each initial handle are created. In the third pass, the tunnel loop is used to generate a refined handle loop. Figure~\ref{fig:torus_initial_pass} depicts these three stages on the simple case of a torus.
Throughout these passes, we mitigate the effect of the random seeding of the diffusion process in the initial pass and we improve the quality of the reported loops. If the focus is only on producing topologically correct loops without any geometric considerations, the final pass may be skipped and the estimates from the initial pass can be used as the handle loops.

\begin{figure}[t]
	\centering
	\includegraphics[width=.24\linewidth]{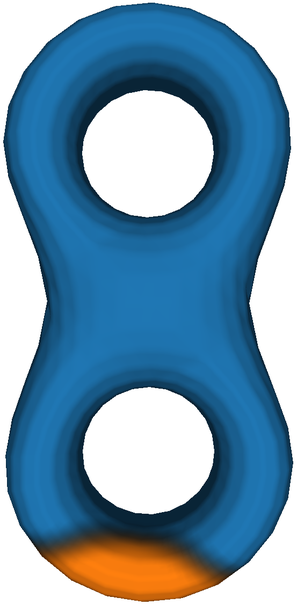}
	\includegraphics[width=.24\linewidth]{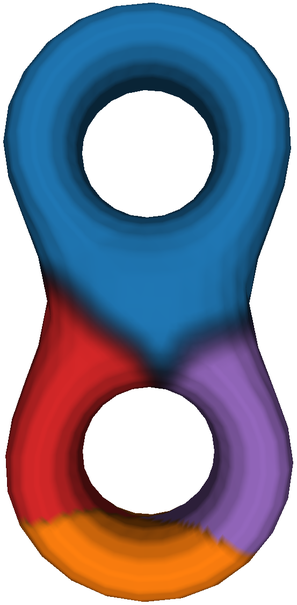}
	\includegraphics[width=.24\linewidth]{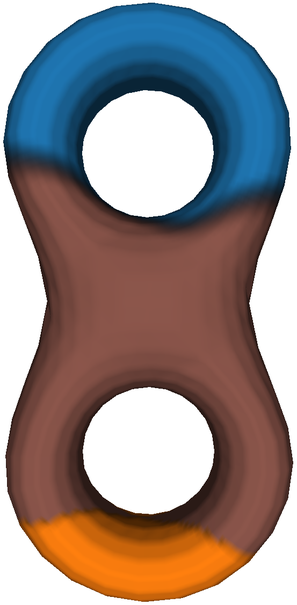}
	\includegraphics[width=.24\linewidth]{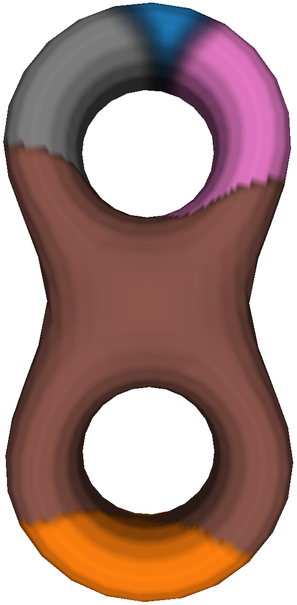}
	\caption{Handle detection on the double torus during the initial pass. The creation of new colors, reflect dynamic changes to multi-layered field representation $\Phi$  such as layer creation or or merging as the advancing fronts split or merge.}
	\label{fig:doubletorus_initial_pass}
\end{figure}

\subsection{Initial Pass}
\label{ssec:initialpass}
The diffusion process is initialized from a random point on the surface and carried out by a simple explicit Euler stepping scheme. After each update, we check the state of the boundary of each advancing front (by inspecting nonzeros along the columns of $\Phi$ which is much cheaper than computing isolines at each iteration. Isolines are computed only when a collision or branching are detected).  We encounter a branching on the surface when there is more than one boundary loop (narrow band), see Figures~\ref{fig:torus_initial_pass}-top-left and~\ref{fig:doubletorus_initial_pass}-left. In this case, each boundary loop is transferred to a new layer (row) in $\Phi$. As $\Phi$ is modeled as a compressed sparse row (CSR) matrix, the transfer means we only have to adjust the column index for each vertex part of a specific connected component.
The parent layer which gave birth to these new fronts can then be ignored in future iterations, because it has no space left to diffuse further, see e.g. Figure~\ref{fig:illustration_loops} for an illustration.
Please note that for visualizations purposes, see Figures,~\ref{fig:torus_initial_pass} and \ref{fig:doubletorus_initial_pass}, the transfer is visualized in a vertex-only coloring and not interpolated, therefore it looks jagged; the transfer itself is fully integrated in the smooth diffusion process.

\begin{figure}[t]
	\centering
	\includegraphics[width=0.49\linewidth]{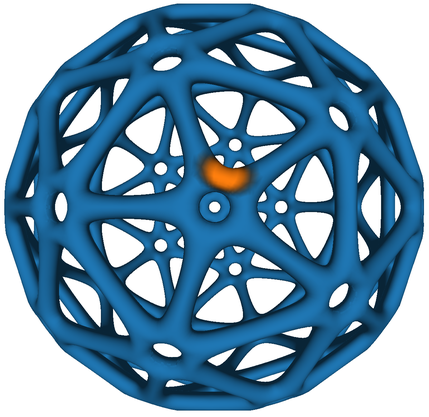}
	\includegraphics[width=0.49\linewidth]{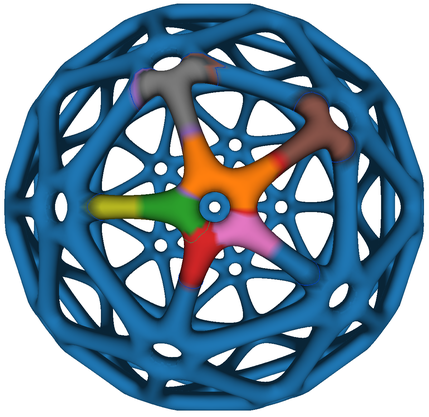}
	\includegraphics[width=0.49\linewidth]{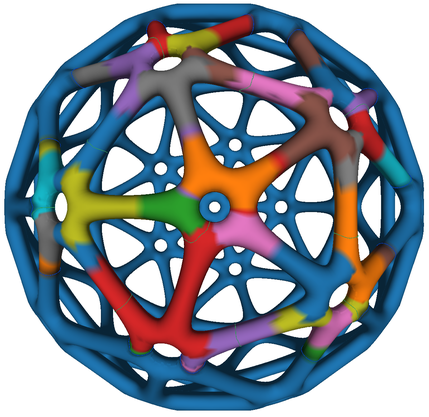}
	\includegraphics[width=0.49\linewidth]{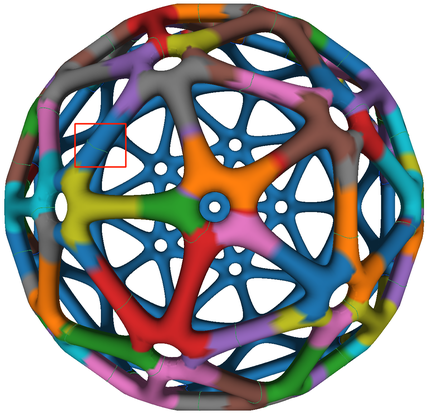}
  \caption{Visualization of the progression of the initial pass on the ball mesh. Note that some handle loops are part of multiple tunnel loops. The red rectangle highlights and example where a naive tunnel loop pass will not report the correct tunnel. The book keeping challenges arising in scenarios like this one highlight the advantage of using a diffuse interface.}
	\label{fig:ball_tunnel_order}
\end{figure}

A handle is detected when multiple advancing fronts meet each other, Figures~\ref{fig:torus_initial_pass}-top-middle and~\ref{fig:doubletorus_initial_pass}-second-left. This can be inferred from $\Phi$ by inspecting which advancing fronts (rows) contribute field values to the same vertices (columns).
Therefore, if a vertex has contributions from two or more layers, those layers should be merged. The number of handles detected is the number of layers involved minus one, e.g., two layers yield one handle.
Inversely, to the scattering of multiple advancing fronts to multiple layers, now we gather all involved layers into a single new layer.
This time the energy values per vertex of all involved layers are accumulated into the energy value of the last involved layer, clearing all the others.
Afterwards the row index of the accumulation result is shifted to represent the new layer.
This process avoids any expensive sparsity pattern changes and is fast to apply.
The handle loops are formed by the involved advancing fronts before they are merged into a single layer, see Figures~\ref{fig:torus_initial_pass}-top-right, ~\ref{fig:doubletorus_initial_pass}-right, and \ref{fig:ball_tunnel_order}.
Since the number of handles equals the number of involved advancing fronts minus one, one boundary loop has to be ignored.
Figure~\ref{fig:torus_initial_pass}-top-right visualizes the formed handle loop as a green line.
In all of our experiments, the longest boundary loop is ignored as this choice seems to give the best visual results. In general, this is not required and our final results are independent of this choice, as long as one layer is ignored.

\subsection{Tunnel loop pass}
\label{ssec:tunnelpass}
It is possible to generate tunnel loops by backtracing streamlines against the diffusion gradient in the initial pass. However this turns out to be a poor algorithmic choice as some tunnel loops may have to pass by the initial random diffusion seed. In order to produce well behaved tunnel loops we take a more judicious approach.
We setup a diffusion process to start on one side of the handle loop whereas the other acts as a border preventing propagation in the opposite side. In this way, the diffusion process can get from one side of the handle loop to other, see Figure~\ref{fig:torus_initial_pass}-middle.
If the other side is reached, we can generate Streamlines back to the start points.
The first point reached on the other side is used as the start point for the streamline trace.
Due to the assumption that the diffusion happens at a constant speed on the surface, this will give us an approximation of the shortest path from one side to the other.
This naive approach, however, does not work for complex geometries, because there may be multiple tunnel candidates for a handle estimate.
The ball shown in figure~\ref{fig:ball_tunnel_order} visualizes that problem.
Whenever a handle estimate lies on a tubular connection between two holes, an example is highlighted by the red rectangle in figure~\ref{fig:ball_tunnel_order}, there will be two possibilities, one on each side, for a tunnel.
Without further restricting the diffusion process we cannot guarantee which tunnel will be found by the Streamlines.
Consequently, the same tunnel may be reported by different handle estimates, if they lie along the same tunnel.
To fix this problem, we introduce an implicit ordering of the reported tunnels.
The second diffusion process for each handle estimate is restricted to the surface area which was already covered at the time the handle estimate was detected.
This ensures that a handle estimate finds the shortest tunnel fully covered by the initial diffusion process and no tunnel is reported more than once.

\begin{figure}[t]
	\centering
	\includegraphics[width=0.32\linewidth]{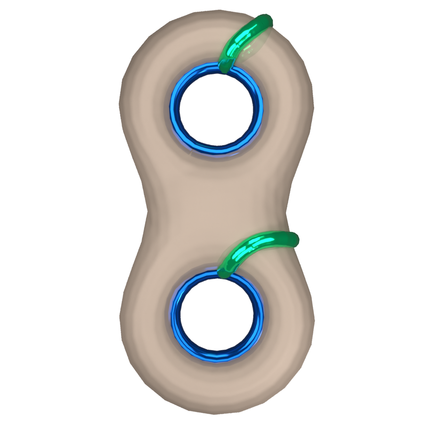}
	\includegraphics[width=0.32\linewidth]{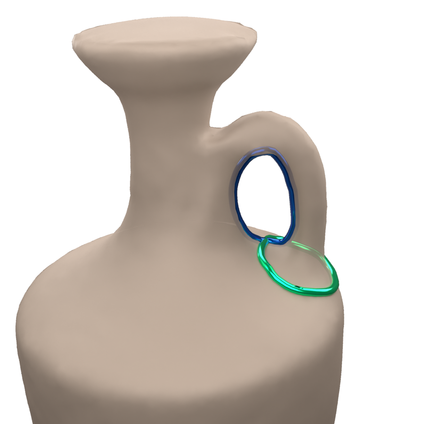}
	\includegraphics[width=0.32\linewidth]{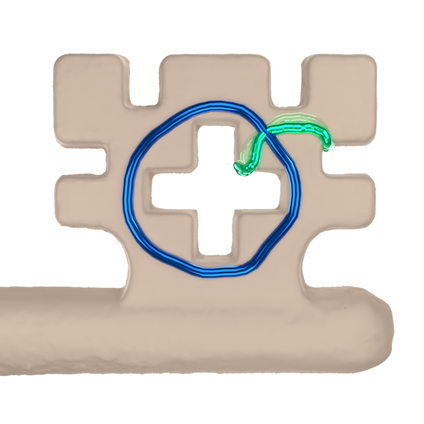}
	\includegraphics[width=0.32\linewidth]{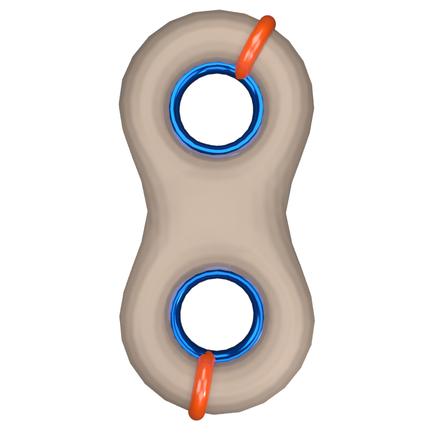}
	\includegraphics[width=0.32\linewidth]{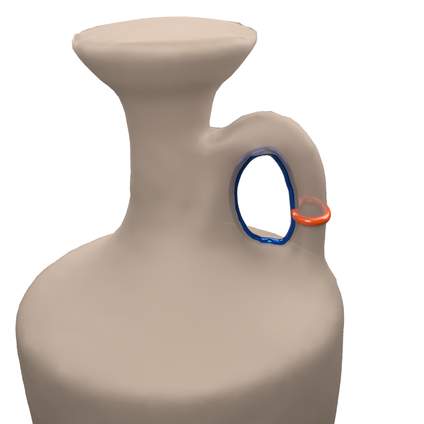}
	\includegraphics[width=0.32\linewidth]{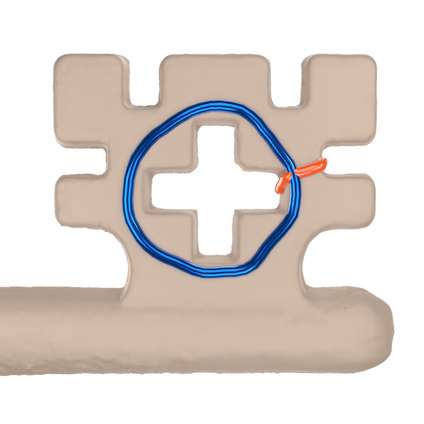}
	\caption{Examples for improved handle loops compared to the estimates of the initial pass. Estimates are rendered as green tubes, tunnel loops are in blue and improved handle loops as orange.}
	\label{fig:handle_loop_improvements}
\end{figure}

\subsection{Handle refinement pass}
\label{ssec:refinepass}
There is a priori no reason for the handle loop generated in the initial pass to be optimally placed or shaped, for instance, perfectly round at the thinnest section of handle. Often times, it will be skewed or twisted, see examples in Figure~\ref{fig:handle_loop_improvements}-top.
By applying the same steps to generate the tunnel loop we can generate a refined handle loop.
We start from one side of the tunnel loop and after we reach the other side, we trace backwards starting from the first point reached, see Figure~\ref{fig:torus_initial_pass}-bottom.
This time the diffusion process is restricted by the corresponding tunnel loop only.
The streamline tracing will yield an improved, ideally placed handle loop, see Figure~\ref{fig:handle_loop_improvements}-bottom. Please note that although our approach is purely heuristic, it is well justified and avoids any costly numerical optimization

\begin{figure}[h]
	\centering
	\includegraphics[width=.35\linewidth]{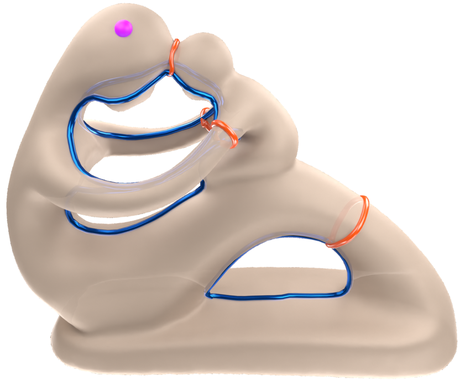}
	\includegraphics[width=.35\linewidth]{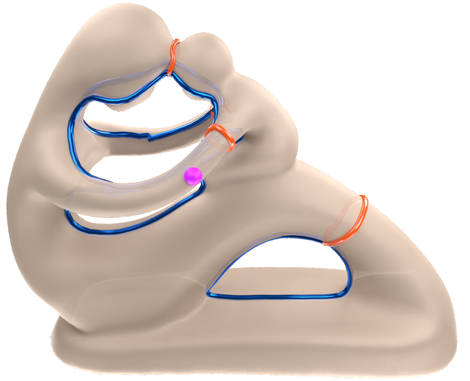}
	\includegraphics[width=.35\linewidth]{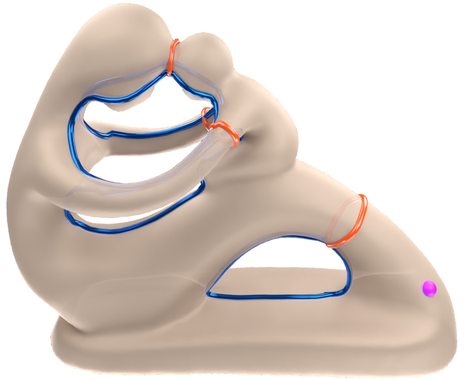}
	\includegraphics[width=.35\linewidth]{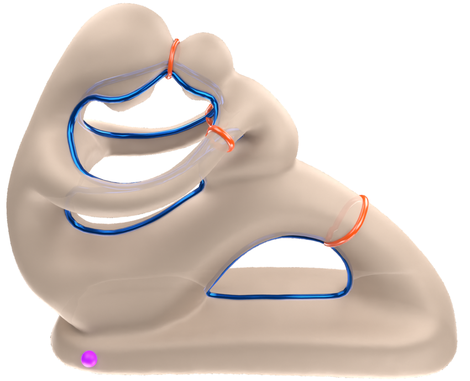}
	\caption{Detected handles and tunnel loops for different initializations. The purple point shows the start point of the diffusion process.}
	\label{fig:different_initial_seeds}
\end{figure}

\subsection{Robustness to initial seed placement}
As discussed above, the multiple pass strategy makes our approach robust to the initial random seed placement. This is confirmed empirically in our test results. A typical scenario is shown in Figure~\ref{fig:different_initial_seeds}. Starting from different locations on the surface of the \emph{mother and child} model, our method generates nearly identical results. Please note that  the handle loops, in orange, slightly differ. This is expected because there are multiple location having similar cross-section thickness alongside the arms. Therefore, there are multiple candidate locations which produce meaningful handle loop.

\begin{figure}[h]
	\centering
	\includegraphics[width=.3\linewidth]{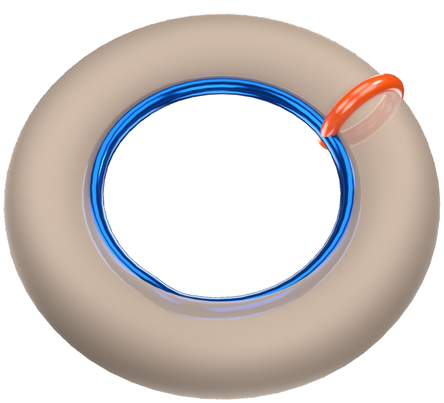}
	\includegraphics[width=.3\linewidth]{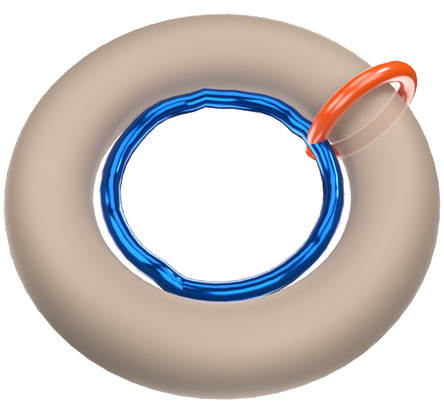}
	\includegraphics[width=.4\linewidth]{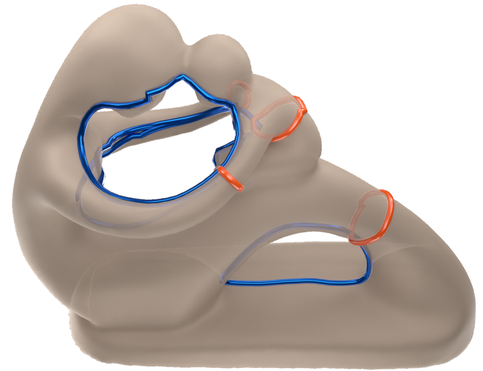}
	\includegraphics[width=.4\linewidth]{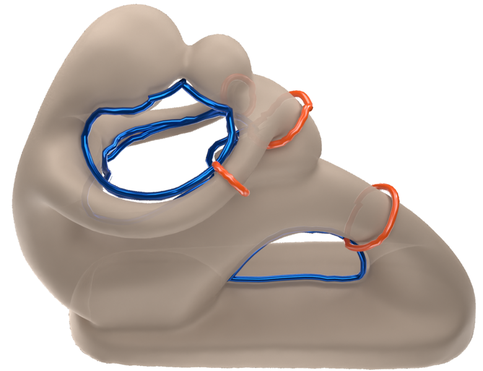}
	\caption{Distinguishing handle and tunnel loops (left) by offsetting them along their corresponding surface normals (right). As illustrated, tunnel loops tend to shrink whereas handle loops expand.}
	\label{fig:normal_adjust}
\end{figure}
\subsection{Distinguishing handle and tunnel loops}
So far, we have assumed that loop handles are found first. This is not always the case,  especially  for deep holes as for instance in the drill in Figure~\ref{fig:result_napoleon}. Our algorithmic flow does not depend on this distinction therefore, we can do identification at the end.  Keeping to the simplicity of our overall approach, basic checks such as slightly offsetting the final loop along its corresponding surface normals and checking the changes in its chord length, as illustrated in Figure~\ref{fig:normal_adjust}, was sufficient for making the distinction in our extensive test cases. Tunnel loops favor hyperbolic parts and will tend to shrink after offsetting whereas handle loops tend to expand. It might be possible to engineer scenarios where this basic distinction fails however we believe this a reasonable price for a tradeoff between fast practical solution and costly bullet proof robustness.

\section{Performance Considerations}

While it would have been easier to implement our current algorithmic pipeline in a multi-threaded fashion, our aim it to harness the tight memory requirement on the gpu and fine grained parallelism pertaining to modern graphics hardware. The two major challenges we faced are finding ways to parallelize the whole algorithm as well as keeping the communication and synchronization between CPU and GPU at a minimum.

As discussed earlier, the time stepping is dominated by Laplacian evaluation which is encoded as sparse matrix matrix multiplication, i.e. SpGEMM, and can readily take advantage of existing numerical kernels.

In all times, our method needs to keep an eye on the evolution of the advancing fronts.  The extraction of the advancing fronts on a layer requires a list of all triangles which are part of the advancing fronts.
A triangle is part of the advancing front, if one or two vertices of that triangle have an energy level above a given threshold and below $1$.
By applying this condition while iterating over all triangles in parallel, we atomically build the list of all triangles part of the advancing front.

Identifying the number of separate advancing fronts, while an easy serial exercise is not straightforward in parallel.  It can then be modeled as a connected component labeling problem on an unstructured grid.
Each marked triangle models a node in the grid and the triangle connectivity describes the edges in the grid.
The connectivity can be extracted from the sparsity pattern of the Laplacian matrix used by the diffusion process.
Connected component labeling is a challenging task to implement efficiently on the GPU.
The nature of this problems requires scattered memory accesses as well as multiple passes, which both are problematic in terms of performance on the GPU.
Our method uses the algorithm described in the work by Soman et al.~\cite{FastCCL}, which minimizes the communication between CPU and GPU and reduces the scattered write operations to improve performance.

\begin{figure}
	\centering
	\includegraphics[width=.15\linewidth]{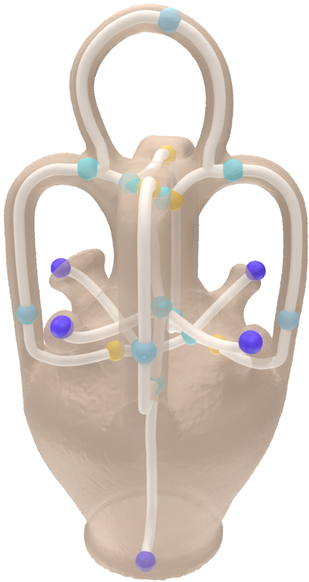}
	\includegraphics[width=.34\linewidth]{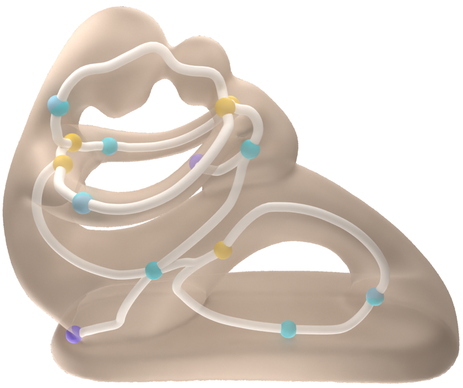}
	\includegraphics[width=.24\linewidth]{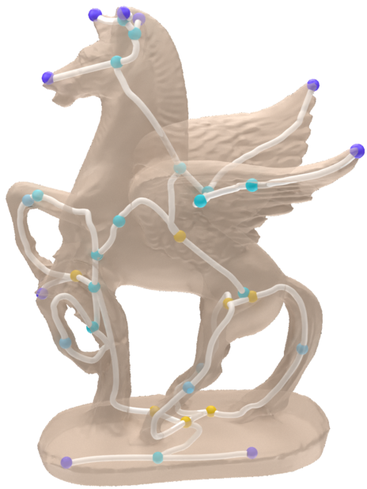}
	\includegraphics[width=.24\linewidth]{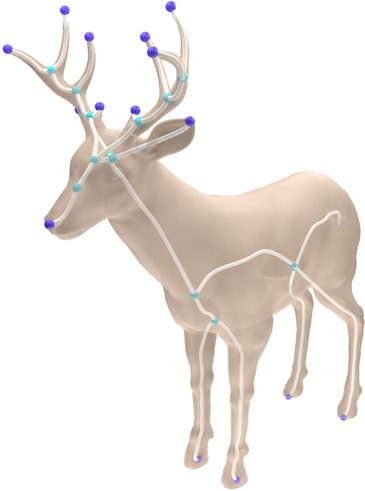}
	\caption{Geometrically embedded Reeb graphs on a selection of relevant test cases.}
	\label{fig:reeb_results}
\end{figure}
\begin{figure}
	\centering
	\includegraphics[width=.2\linewidth]{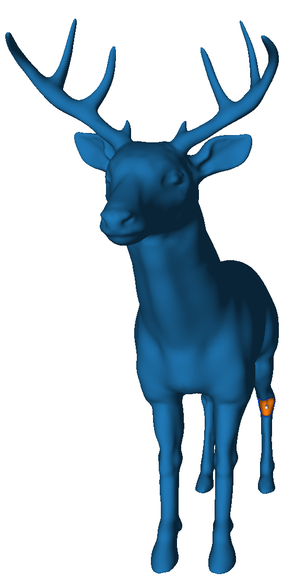}
	\includegraphics[width=.2\linewidth]{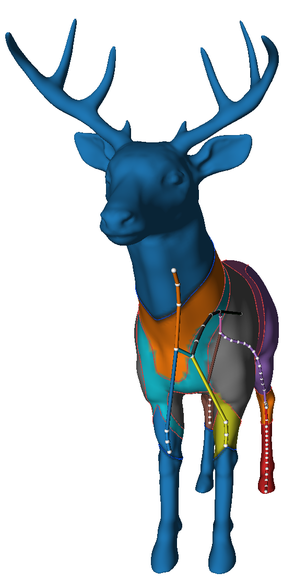}
	\includegraphics[width=.2\linewidth]{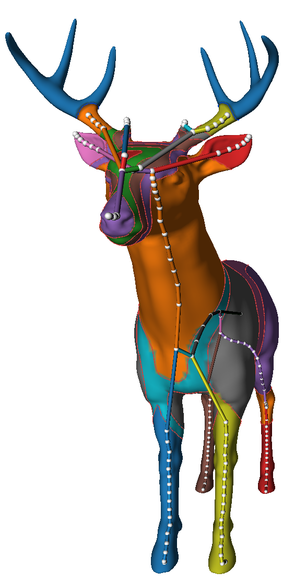}
	\includegraphics[width=.2\linewidth]{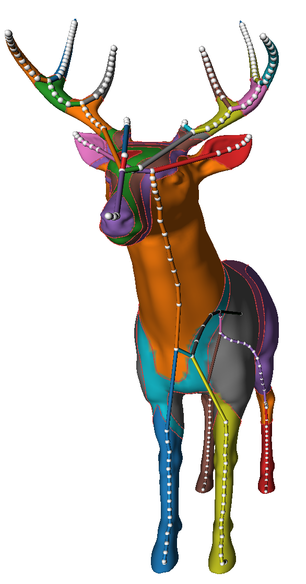}
	\caption{Reeb graph creation on the genus-$0$ deer model}
	\label{fig:reeb_deer}
\end{figure}

The	tunnel loop pass (subsection~\ref{ssec:tunnelpass}) and handle refinement pass (subsection~\ref{ssec:refinepass}) only differ in the initialization of the diffusion process and are discussed as one.
Compared to the diffusion process during the initial pass, in these passes, the diffusion process has exactly one single layer.
No scattering or merging is performed.
This allows us to replace the sparse resizable system matrix $\Phi$ by a dense vector.
The dense vector contains the energy value for each vertex on the single layer.
The base layer, required by the diffusion process, is implicitly modeled as one minus the energy value of the first layer.
Furthermore, the SpGEMM during the diffusion update is replaced by a simple sparse matrix vector multiplication (SpMV).
This performance optimization of the diffusion process is particularly important because each tunnel and handle loop requires a separate diffusion process.
To fully utilize the GPU, multiple gradient and handle refinement passes run in parallel.
Due to the constant memory consumption by the dense vector, we can calculate the exact number of possible parallel passes based on the available GPU memory.

\begin{figure*}[h]
	\centering
	\includegraphics[width=.18\linewidth]{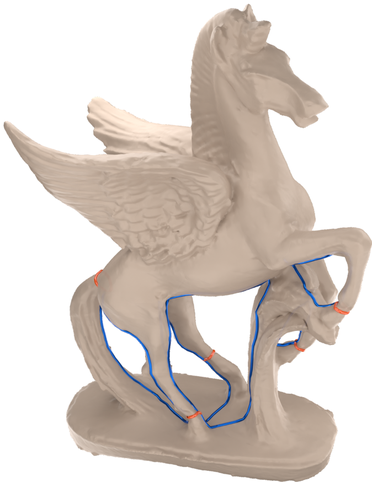}
	\includegraphics[width=.3\linewidth]{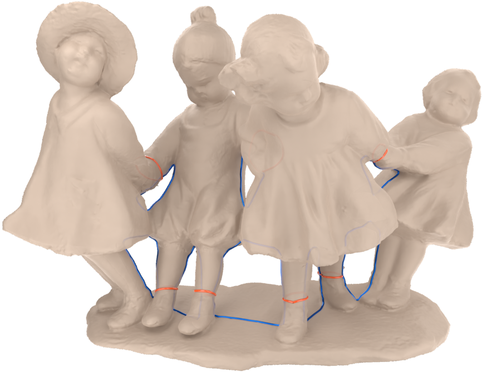}
	\includegraphics[width=.14\linewidth]{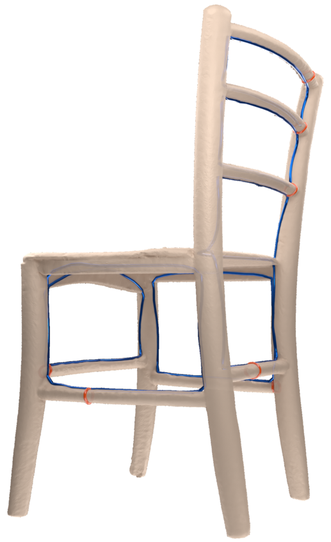}
	\includegraphics[width=.15\linewidth]{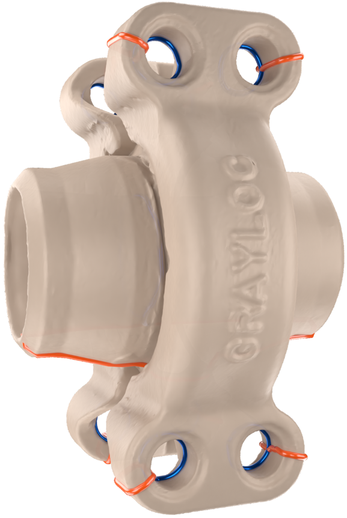}
	\includegraphics[width=.1\linewidth]{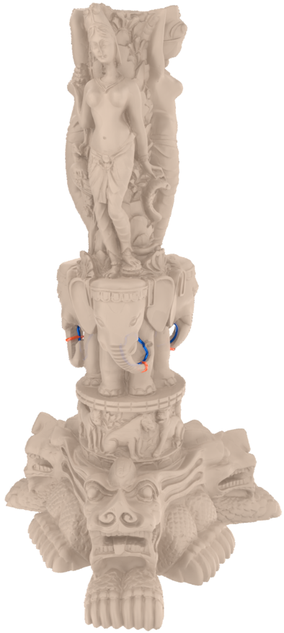}
	\includegraphics[width=.23\linewidth]{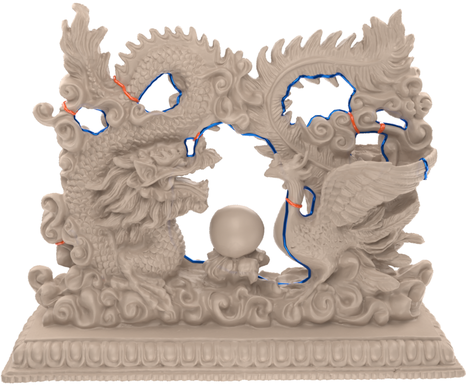}
	\includegraphics[width=.32\linewidth]{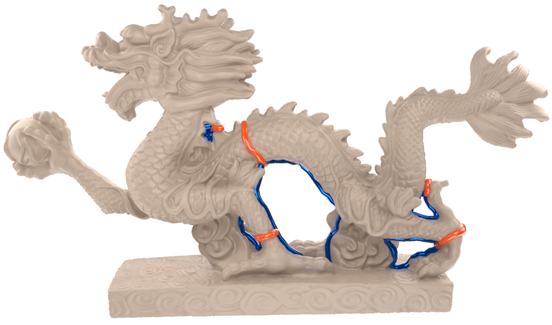}
	\includegraphics[width=.07\linewidth]{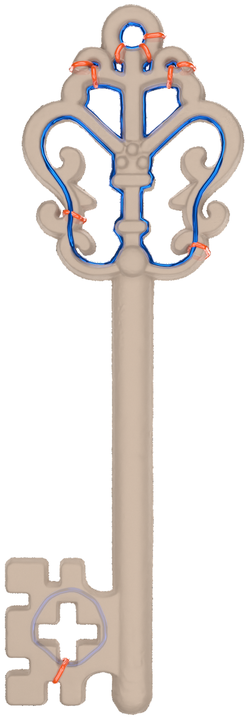}
	\includegraphics[width=.22\linewidth]{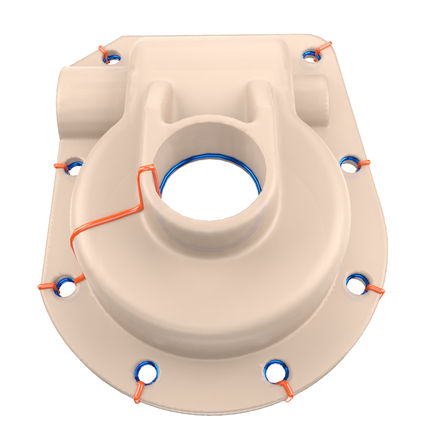}
	\includegraphics[width=.1\linewidth]{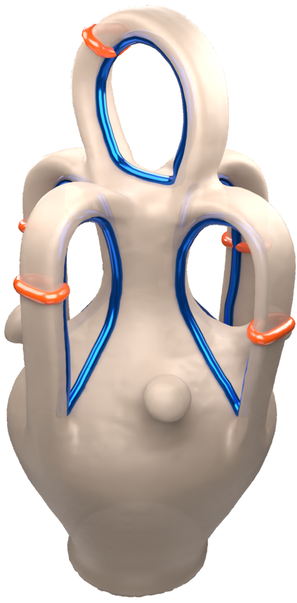}
	\caption{Typical results of our approach on multiple data sets: From top left, The Pegasus, dancing children, chair, Grayloc, Thai statue, Dragon Phoenix, Dragon Ball, Metal Key, Casting, and Botijo Jar.  }
	\label{fig:results_set}
\end{figure*}

\section{Reeb Graph Extraction}

While our objective has been the detection of handle tunnel and loops, looking at Figure~\ref{fig:doubletorus_initial_pass} it is clear that the diffusion process somewhat capture all the branching in the surface. While the topological information encoded by a function defined on given surface by tracking the connected components of it level sets is generally captured in the notion of Reeb graphs. Interest into Reeb graphs in computer graphics goes back to the work of \cite{ReebAlgoIntroduction} which opened the door for a steady research effort on the topic. This includes performance oriented solutions, e.g.~\cite{Pascucci2007}, mainstream implementations such as the on available from within the \emph{Topology Toolkit}~\cite{ttk} , improvements to the algorithmic complexity, eg.~\cite{Doraiswamy2009}, and extensions and generalizations e.g.~\cite{Biasotti2000}.

In this digression, our goal is not produce a state of the art Reeb graph algorithm but to show the versatility of our approach, in the sense, that by the same token we get the relevant handle and tunnel loops and meaningful geometric embedding of the Reeb graph associated with our diffusion on the surface. 

Just using the diffusion initialized at random point as described in the subsection~\ref{ssec:initialpass} would naturally yield topologically correct graph, however as we seek to associate a geometric embedding to the surface, we use the second pass from subsection \ref{ssec:tunnelpass} to improve the location of the junctions. Typical results of our approach are shown in Figure~\ref{fig:reeb_results}. Clearly for surfaces without any handle or tunnel loops it is sufficient to run the initial pass to track all the relevant branching, as illustrated in the case of the deer model in Figure~\ref{fig:reeb_deer} . The location of the connecting points across the geometric embedding are obtained by simply taking the mean of the narrow band. 

An estimate of the performance can be inferred by comparing the cost of (initial and secondary pass) from table \ref{tab:step_disp} to one of the top performing methods~\cite{Pascucci2007} since it is the method used in the Reeb graph step in~\cite{dey2013}.
\begin{scriptsize}
\begin{table*}[h]
\centering
\begin{tabular}{@{\extracolsep{4pt}}lcccccccccccc}
  \toprule
		 & \multicolumn{4} {c} {Ours}       &\multicolumn{5} {c} {\cite{dey2013}} & Length ratio \\   \cmidrule{2-5}\cmidrule{6-10}\cmidrule{11-11}
  Mesh(\#faces, Genus) &Total	&Initial Pass	&Tunnel Pass	&Handle Pass	&Total	&Reeb graph	&Map/Link	&Edge annot.	&Shorten. & ours/\cite{dey2013} \\ \cmidrule{2-5}  \cmidrule{6-10} \cmidrule{11-11}
	Torus	(3.1k, 1) &0.16	&0.12	&0.00	&0.00	&0.02	&0.01	&0	&0	&0.01 	&1.005\\
	Double Torus	(25k, 2) &0.48	&0.42	&0.00	&0.00	&0.18	&0.05	&0.01	&0.02	&0.1 		&1.010\\
	MotherWithChild(28K,4)	&0.42	&0.29	&0.08	&0.04	&0.52	&0.05	&0.05	&0.02	&0.40	&0.988\\
	Botijo(82K,5)	&1.78	&0.65	&0.72	&0.41	&1.81	&0.21	&0.17	&0.11	&1.32 	&0.960\\
	Casting(93K,9)			&1.31	&0.70	&0.30	&0.30	&3.10	&0.40	&0.10	&0.16	&2.44 	&0.999\\
	Happy Buddha(98K,8)	&1.13	&0.61	&0.28	&0.24	&2.04	&0.24	&0.10	&0.15	&1.55 	&0.988\\
	Ball	(184K, 120)	&17.65	&13.01	&1.97	&1.16	&401.77	&0.28	&34.76	&3.16	&363.57	&0.918\\
	Metal Key(390k,10)		&10.48	&6.87	&2.11	&1.50	&36.52	&10.35	&9.52	&0.52	&16.13	&0.984\\
	Wooden Chair(400K,7)	&7.49	&4.45	&2.28	&0.77	&31.21	&2.77	&18.55	&0.43	&9.46 	&0.982\\
	Dragon Phoenix (750K, 11) &12.89	&8.75	&2.15	&1.99	&100.83	&9.77	&50.66	&1.73	&38.67	&0.942\\
	Grayloc	(921k, 5) &22.25	&14.23	&4.11	&3.92 &DNF & & & &\\
	V745 Sco nova (923K, 184) & 273.99 &104.21 & 86.32 & 82.34 & DNF & & & &\\
	Metal Table(950K, 198)	&151.33	&92.31	&36.02	&21.32	&3,732.92	&4.49	&35.74	&16.68	&3,676.01		&0.954\\
	Dancing children(1.3M,8)  	&20.57	&13.29	&4.01	&3.13	&102.54	&38.26	&7.60	&3.16	&53.52 	&0.989\\
	Drill	(1.5m, 13) &18.77	&13.92	&2.67	&2.18	&90.32	&13.52	&1.14	&2.74	&72.92	&0.846\\
	Dragon Tamer(2M, 335)	&429.57	&254.93	&93.90	&78.63	&>20,000.00	&16.47	&885.50	&249.81	&Crash	&\\
  	Dragon Ball (2.4M, 5, 40) &37.29	&21.88	&9.40	&6.01	&91.70	&38.14	&1.28	&2.84	&49.44	&0.975\\
	Napoleon	(6.5M, 25) &705.34	&376.50	&170.86		&157.99		&>2,970.16	&2730.67	&239.49	&Crash	&\\
	Statue	(10M, 3) &167.71	&152.18	&7.41	&8.12	&491.69	&253.45	&19.12	&7.91	&211.21	&0.976\\
	\bottomrule
\end{tabular}
	\caption{Runtime comparison for various test models. Timings are measured in seconds. The length ratio compares the total handle and tunnel loops length of our method against~\cite{dey2013}.}
\label{tab:step_disp}
\end{table*}
\end{scriptsize}

\begin{tiny}
\begin{table}
\centering
\begin{tabular}{@{\extracolsep{4pt}}lcccc}
  \toprule
  & \multicolumn{3} {c} {Ours} & \cite{dey2013} \\ \cmidrule{2-4}\cmidrule{5-5}
  Mesh(\#faces, Genus, Mem. Size) & Inital Pass & T/H. Pass & T/H. Pass (\#threads) & Total \\\cmidrule{2-4}\cmidrule{5-5}
  Torus	(3.1k, 1, 0.05) &86	&2	&2 (1)	&10\\
  Double Torus	(25k, 2, 0.4) &92	&4	&6 (2)	&42\\
  MotherWithChild (28K, 4, 0.48)	&92	&6	&18 (4)	&76\\
  Botijo (82K, 5, 1.5)	&108	&64	&108 (5)	&193\\
  Casting(93K,9, 1.6)			&114	&18	&134 (9)	&210\\
  Happy Buddha(98K,8,1.8)	&97	&16	&112 (8)	&189\\
  Ball	(184K, 120, 3.15)	&152	&152	&458 (10)	&2135\\
  Metal Key (390k,10, 6.6)		&210	&82	&552 (10)	&810\\
  Wooden Chair (400K,7, 6.8)	&214	&82	&472 (7)	&921\\
  Dragon Phoenix (750K, 11, 12) &340	&106	&1062 (10)	&1540\\
  Grayloc	(921k, 5, 16) &366	&354	&1155 (5)	&0\\
  V745 Sco nova (923K, 184, 16)	&502	&187	&1402 (10)	& DNF\\
  Metal Table (950K,198, 16)	&336	&166	&882 (10)	&16891\\
  Dancing children (1.3M,8, 21)  	&462	&240	&612 (8)	&4992\\
  Drill	 (1.5m, 13, 25) &496	&338	&1844 (13)	&4142\\
  Dragon Tamer (2M, 335, 34)	&636	&290	&1410 (10)	&>32562\\
  Dragon Ball (2.4M, 5, 40)   &728	&292	&1452 (5)	&4557 \\
  Napoleon	(6.5M, 25, 111) &1844	&1566	&7744 (10)	&>15000\\
  Statue	(10M, 3, 171) &496	&338	&1844 (13)	&4142\\
  \bottomrule
\end{tabular}
\caption{Memory consumption for various test models. For our method peak GPU memory usage is reported for the initial pass and the tunnel and handle (H./T.) pass combined.
          For the tunnel and handle passes, we report the memory consumption with and without running multiple tunnel/handle pairs in parallel.
		  The initial pass and the gradients pass do run one after another, therefore, their maximum is the maximum of our whole method.
          For \cite{dey2013} peak CPU memory consumption is reported.
          }
\label{tab:memory}
\end{table}
\end{tiny}

\section{Results}
\label{sec:results}

All our experiment were carried on a Intel(R) Core(TM) i7-7700 with  32Gb system memory and a Nvidia RTX 2080ti 11gb as GPU.
As initialization point for the initial diffusion process we used the first vertex of the mesh.

Figure \ref{fig:results_set} shows a collection of non-trivial test case meshes featuring: large loops as in the pegasus model and the dancing children, object with thin features such as the key and the chairs and the mechanical parts. In all these cases, our approaches detect the handles and tunnels correctly.
The handles and tunnels in the dragon models are heavily decorated with ornaments. The reported tunnel loops correctly avoid following the ornaments which face inwards compared to the tunnel loop. Following those ornaments would increase the length of the loop. The handle loops are located at the smallest parts despite the ornaments.

\begin{figure*}[h]
	\centering
	\includegraphics[width=0.42\linewidth]{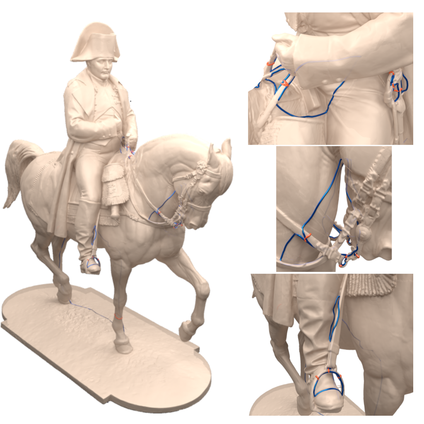}
	\includegraphics[width=0.27\linewidth]{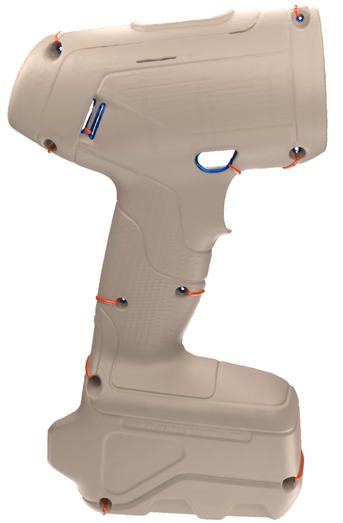}
	\includegraphics[width=0.28\linewidth]{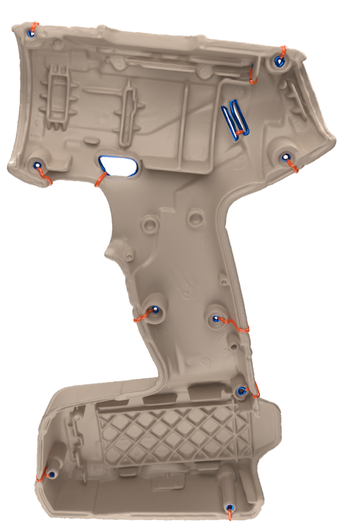}
	\caption{Evaluation of our approach on the Napoleon model (with zoomed views) and the Drill model (double sided view). }
	\label{fig:result_napoleon}
\end{figure*}

Figure \ref{fig:result_napoleon} shows the tunnel and handles loops computed for the napoleon mesh.
This mesh features very large tunnels, e.g. the tunnel covering the stone plate, two legs and the body of the horse, and also very tiny ones, e.g. at the top part of the foot rest or at the reins.

The drill shown in Figure \ref{fig:result_napoleon} features very small tunnels combined with large handles.
All tunnels and handles are reported correctly, because our method is not based on any presumptions on the length of handle or tunnel loops.
Note that not all screw holes feature a tunnel in the mesh, since it is a 3D scan.

\begin{figure}
	\centering
	\includegraphics[width=0.58\linewidth]{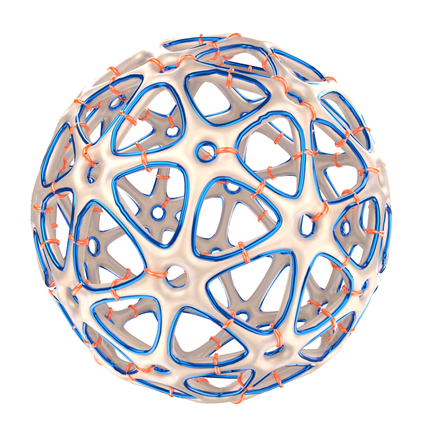}
	\includegraphics[width=.4\linewidth]{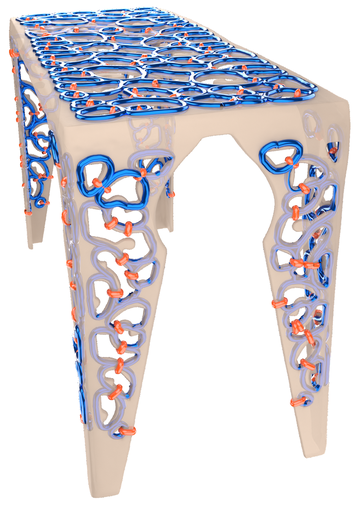}
	\includegraphics[width=.85\linewidth]{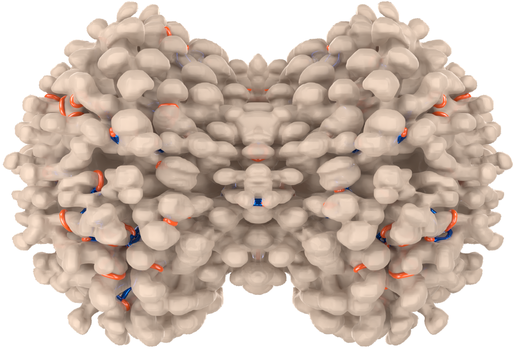}

	\caption{Very high genus test cases, a complicated ball, a metal table, and  a model of the V745 Sco nova (courtesy of NASA). }
	\label{fig:result_table}
\end{figure}

Table~\ref{tab:step_disp} compares the runtime of our method with the state of the art approach~\cite{dey2013}.
The results show that our method is slower for relatively small meshes but scales significantly better with larger meshes.
This can be explained by the constant overhead of GPU initialization as well as memory transfers to and from the GPU to obtain the results.
Our experiments included five meshes with a high genus: ball, metal table, V745 Sco nova and dragon tamer.
The results of these meshes are visualized in figure~\ref{fig:result_table} and figure~\ref{fig:teaser}.
In these cases our method outperformed the other method by over an order of magnitude.
V745 Sco nova and dragon tamer, which has the highest genus of all tested meshes, did not complete with their method.
Also the metal table required multiple runs until the method completed without an error, we report only the runtime of the successful run.

The number of pairs detected by both methods is the same. As it is cumbersome to match the pairs of both methods automatically for comparison, we compare the total length of all loops. The length ratio in Table~\ref{tab:step_disp}  suggest that our method yields shorter loops in general. Please not that the torus and the double torus are regularly meshed models where the optimal tunnel and handles loops are well aligned with the mesh edges. Since the method of ~\cite{dey2013} performs the shortening in combinatorial fashion it lands directly on those edges. Still the difference with our method is not significant.

Table~\ref{tab:memory} contains the memory consumptions obtained during our experiments.
Again, except for very small meshes, our method uses less memory even when multiple handle and tunnel refinement passes run in parallel.
We report the minimum memory requirement for the handle and tunnel pass as well as the maximum when running multiple passes in parallel.
The memory consumption does not directly scale by the number of parallel runs because some read only information, such as the mesh, can be shared between these passes.
Our initial tests showed that running more than 10 passes in parallel did not improve the performance further.
This is probably caused by driver overhead scheduling the parallel kernel runs on the GPU.

\paragraph{Limitations and discussion.}
The performance of  \cite{dey2013} depends largely on the fast Reeb graph computations of ~\cite{Pascucci2007} which in turn uses “free” height functions (coordinates) as Morse functions. Certainly, other functions can be used, the “streaming meshes” format has been used in the latter but then the cost of the underlying Fiedler vector computation should be factored in. Similarly, when geodesic or harmonic functions, e.g.,~\cite{Hilaga} are used, their cost will impede performance.  Therefore the comparisons we are providing are representative.

Our approach operates on the piecewise linear manifold setting, and cannot offer the guarantees of well and long-established combinatorial approaches but empirical evidence shows that our method produces the same number of pairs as [Dey2013]. This is expected since the method relies on first principals in topology embodied in the loop shrinking property. Our method builds on the ability to conduct standard numerical simulations on surface meshes and is therefore bound by fairly well-known requirements on methods such as the finite element method. We do not see this as a limitation but rather a motivation for exploring common grounds between combinatorial and variational approaches.

\begin{figure}[h]
	\centering

	\includegraphics[width=.3\linewidth]{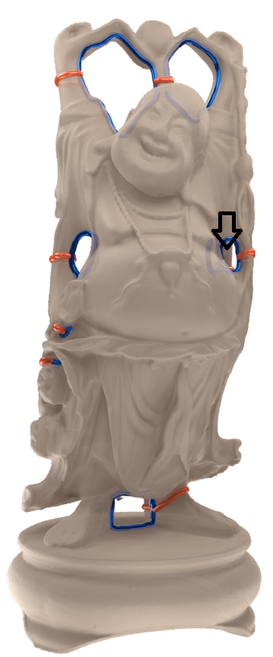}
	\includegraphics[width=.65\linewidth]{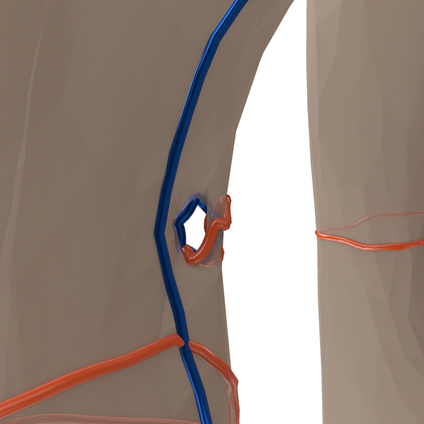}
	\caption{The Happy Buddha model (left) and a zoom-in on the area marked by the arrow showing the effect of a tiny coarse handle on the quality of our result.}
	\label{fig:buddha}
\end{figure}

There can be scenarios where tiny and coarsely meshed handles can affect the quality of our results or even escape the width of our advancing front.  The only test case where we encountered such a scenario is depicted in Figure \ref{fig:buddha}. Although the handle is only spanned by three triangles, our method still captures it but the loop is clearly suboptimal. Although this does not respect the smoothness assumptions of the linear manifold required in our variational setting, it can be addressed by adaptive refinement along the front.
\section{Conclusion}

We have re-abstracted the shrinking loop property as a continuous growth process steered by diffusion.
The diffusion field was modeled using diffuse interface and the dynamics of the advancing fronts were captured by assuming a multilayered field representation where the creation and and merging of layers are respectively steered by front splitting and collision.
Our approach is simple and versatile allowing the detection of handle and tunnel loops as well as the creation of Reeb graphs. Furthermore the time stepping nature of our approach allows for a close control of the whole process by simple adjustment of the field and regions of diffusion. We foresee that this control will help extend our approach to other application such topological surgery, mesh segmentation, and rigging. We hope that adopting our methodology will help break the deadlock in performance and bring new machinery to help understanding and harnessing geometric problems.


\bibliographystyle{ACM-Reference-Format}
\bibliography{references}


\begin{thebibliography}{26}


\ifx \showCODEN    \undefined \def \showCODEN     #1{\unskip}     \fi
\ifx \showDOI      \undefined \def \showDOI       #1{#1}\fi
\ifx \showISBNx    \undefined \def \showISBNx     #1{\unskip}     \fi
\ifx \showISBNxiii \undefined \def \showISBNxiii  #1{\unskip}     \fi
\ifx \showISSN     \undefined \def \showISSN      #1{\unskip}     \fi
\ifx \showLCCN     \undefined \def \showLCCN      #1{\unskip}     \fi
\ifx \shownote     \undefined \def \shownote      #1{#1}          \fi
\ifx \showarticletitle \undefined \def \showarticletitle #1{#1}   \fi
\ifx \showURL      \undefined \def \showURL       {\relax}        \fi
\providecommand\bibfield[2]{#2}
\providecommand\bibinfo[2]{#2}
\providecommand\natexlab[1]{#1}
\providecommand\showeprint[2][]{arXiv:#2}

\bibitem[\protect\citeauthoryear{Biasotti, Falcidieno, and Spagnuolo}{Biasotti
  et~al\mbox{.}}{2000}]%
        {Biasotti2000}
\bibfield{author}{\bibinfo{person}{Silvia Biasotti}, \bibinfo{person}{Bianca
  Falcidieno}, {and} \bibinfo{person}{Michela Spagnuolo}.}
  \bibinfo{year}{2000}\natexlab{}.
\newblock \showarticletitle{Extended Reeb Graphs for Surface Understanding and
  Description}. In \bibinfo{booktitle}{\emph{Discrete Geometry for Computer
  Imagery}}, \bibfield{editor}{\bibinfo{person}{Gunilla Borgefors},
  \bibinfo{person}{Ingela Nystr{\"o}m}, {and}
  \bibinfo{person}{Gabriella~Sanniti di~Baja}} (Eds.).
  \bibinfo{publisher}{Springer Berlin Heidelberg}, \bibinfo{address}{Berlin,
  Heidelberg}, \bibinfo{pages}{185--197}.
\newblock
\showISBNx{978-3-540-44438-1}


\bibitem[\protect\citeauthoryear{Brezovsky, Chovancova, Gora, Pavelka,
  Biedermannova, and Damborsky}{Brezovsky et~al\mbox{.}}{2013}]%
        {brezovsky2013software}
\bibfield{author}{\bibinfo{person}{Jan Brezovsky}, \bibinfo{person}{Eva
  Chovancova}, \bibinfo{person}{Artur Gora}, \bibinfo{person}{Antonin Pavelka},
  \bibinfo{person}{Lada Biedermannova}, {and} \bibinfo{person}{Jiri
  Damborsky}.} \bibinfo{year}{2013}\natexlab{}.
\newblock \showarticletitle{Software tools for identification, visualization
  and analysis of protein tunnels and channels}.
\newblock \bibinfo{journal}{\emph{Biotechnology advances}}
  \bibinfo{volume}{31}, \bibinfo{number}{1} (\bibinfo{year}{2013}),
  \bibinfo{pages}{38--49}.
\newblock


\bibitem[\protect\citeauthoryear{Chen, Jester, and Gopi}{Chen
  et~al\mbox{.}}{2018}]%
        {Chen2018}
\bibfield{author}{\bibinfo{person}{Jia Chen}, \bibinfo{person}{James Jester},
  {and} \bibinfo{person}{M. Gopi}.} \bibinfo{year}{2018}\natexlab{}.
\newblock \showarticletitle{Fast Computation of Tunnels in Corneal Collagen
  Structure}. In \bibinfo{booktitle}{\emph{Proceedings of Computer Graphics
  International 2018}} \emph{(\bibinfo{series}{CGI 2018})}.
  \bibinfo{publisher}{Association for Computing Machinery},
  \bibinfo{address}{New York, NY, USA}, \bibinfo{pages}{57–65}.
\newblock
\showISBNx{9781450364010}
\urldef\tempurl%
\url{https://doi.org/10.1145/3208159.3208175}
\showDOI{\tempurl}


\bibitem[\protect\citeauthoryear{de~Verdi\`{e}re and Erickson}{de~Verdi\`{e}re
  and Erickson}{2006}]%
        {deVerdiere2006}
\bibfield{author}{\bibinfo{person}{\'{E}ric~Colin de Verdi\`{e}re} {and}
  \bibinfo{person}{Jeff Erickson}.} \bibinfo{year}{2006}\natexlab{}.
\newblock \showarticletitle{Tightening Non-Simple Paths and Cycles on
  Surfaces}. In \bibinfo{booktitle}{\emph{Proceedings of the Seventeenth Annual
  ACM-SIAM Symposium on Discrete Algorithm}} \emph{(\bibinfo{series}{SODA
  '06})}. \bibinfo{publisher}{Society for Industrial and Applied Mathematics},
  \bibinfo{address}{USA}, \bibinfo{pages}{192–201}.
\newblock
\showISBNx{0898716055}


\bibitem[\protect\citeauthoryear{Dey, Fan, and Wang}{Dey et~al\mbox{.}}{2013}]%
        {dey2013}
\bibfield{author}{\bibinfo{person}{Tamal~K. Dey}, \bibinfo{person}{Fengtao
  Fan}, {and} \bibinfo{person}{Yusu Wang}.} \bibinfo{year}{2013}\natexlab{}.
\newblock \showarticletitle{An Efficient Computation of Handle and Tunnel Loops
  via Reeb Graphs}.
\newblock \bibinfo{journal}{\emph{ACM Trans. Graph.}} \bibinfo{volume}{32},
  \bibinfo{number}{4}, Article \bibinfo{articleno}{32} (\bibinfo{date}{July}
  \bibinfo{year}{2013}), \bibinfo{numpages}{10}~pages.
\newblock
\showISSN{0730-0301}
\urldef\tempurl%
\url{https://doi.org/10.1145/2461912.2462017}
\showDOI{\tempurl}


\bibitem[\protect\citeauthoryear{Dey, Li, and Sun}{Dey et~al\mbox{.}}{2007}]%
        {dey2007}
\bibfield{author}{\bibinfo{person}{Tamal~K. Dey}, \bibinfo{person}{Kuiyu Li},
  {and} \bibinfo{person}{Jian Sun}.} \bibinfo{year}{2007}\natexlab{}.
\newblock \showarticletitle{On Computing Handle and Tunnel Loops}. In
  \bibinfo{booktitle}{\emph{Proceedings of the 2007 International Conference on
  Cyberworlds}} \emph{(\bibinfo{series}{CW '07})}. \bibinfo{publisher}{IEEE
  Computer Society}, \bibinfo{address}{USA}, \bibinfo{pages}{357–366}.
\newblock
\showISBNx{0769530052}


\bibitem[\protect\citeauthoryear{Dey, Li, Sun, and Cohen-Steiner}{Dey
  et~al\mbox{.}}{2008}]%
        {Dey2008}
\bibfield{author}{\bibinfo{person}{Tamal~K. Dey}, \bibinfo{person}{Kuiyu Li},
  \bibinfo{person}{Jian Sun}, {and} \bibinfo{person}{David Cohen-Steiner}.}
  \bibinfo{year}{2008}\natexlab{}.
\newblock \showarticletitle{Computing Geometry-Aware Handle and Tunnel Loops in
  3D Models}.
\newblock \bibinfo{journal}{\emph{ACM Trans. Graph.}} \bibinfo{volume}{27},
  \bibinfo{number}{3} (\bibinfo{date}{Aug.} \bibinfo{year}{2008}),
  \bibinfo{pages}{1–9}.
\newblock
\showISSN{0730-0301}
\urldef\tempurl%
\url{https://doi.org/10.1145/1360612.1360644}
\showDOI{\tempurl}


\bibitem[\protect\citeauthoryear{Diaz-Gutierrez, Eppstein, and
  Gopi}{Diaz-Gutierrez et~al\mbox{.}}{2009}]%
        {Diaz-Gutierrez2009}
\bibfield{author}{\bibinfo{person}{P. Diaz-Gutierrez}, \bibinfo{person}{D.
  Eppstein}, {and} \bibinfo{person}{M. Gopi}.} \bibinfo{year}{2009}\natexlab{}.
\newblock \showarticletitle{{Curvature Aware Fundamental Cycles}}.
\newblock \bibinfo{journal}{\emph{Computer Graphics Forum}}
  (\bibinfo{year}{2009}).
\newblock
\showISSN{1467-8659}
\urldef\tempurl%
\url{https://doi.org/10.1111/j.1467-8659.2009.01580.x}
\showDOI{\tempurl}


\bibitem[\protect\citeauthoryear{Doraiswamy and Natarajan}{Doraiswamy and
  Natarajan}{2009}]%
        {Doraiswamy2009}
\bibfield{author}{\bibinfo{person}{Harish Doraiswamy} {and}
  \bibinfo{person}{Vijay Natarajan}.} \bibinfo{year}{2009}\natexlab{}.
\newblock \showarticletitle{Efficient Algorithms for Computing Reeb Graphs}.
\newblock \bibinfo{journal}{\emph{Comput. Geom. Theory Appl.}}
  \bibinfo{volume}{42}, \bibinfo{number}{6–7} (\bibinfo{date}{Aug.}
  \bibinfo{year}{2009}), \bibinfo{pages}{606–616}.
\newblock
\showISSN{0925-7721}
\urldef\tempurl%
\url{https://doi.org/10.1016/j.comgeo.2008.12.003}
\showDOI{\tempurl}


\bibitem[\protect\citeauthoryear{El-Sana and Varshney}{El-Sana and
  Varshney}{1997}]%
        {elsana1997}
\bibfield{author}{\bibinfo{person}{Jihad El-Sana} {and}
  \bibinfo{person}{Amitabh Varshney}.} \bibinfo{year}{1997}\natexlab{}.
\newblock \showarticletitle{Controlled Simplification of Genus for Polygonal
  Models}. In \bibinfo{booktitle}{\emph{Proceedings of the 8th Conference on
  Visualization '97}} \emph{(\bibinfo{series}{VIS '97})}.
  \bibinfo{publisher}{IEEE Computer Society Press},
  \bibinfo{address}{Washington, DC, USA}, \bibinfo{pages}{403–ff.}
\newblock
\showISBNx{1581130112}


\bibitem[\protect\citeauthoryear{Eppstein}{Eppstein}{2003}]%
        {Eppstein2003}
\bibfield{author}{\bibinfo{person}{David Eppstein}.}
  \bibinfo{year}{2003}\natexlab{}.
\newblock \showarticletitle{Dynamic Generators of Topologically Embedded
  Graphs}. In \bibinfo{booktitle}{\emph{Proceedings of the Fourteenth Annual
  ACM-SIAM Symposium on Discrete Algorithms}} \emph{(\bibinfo{series}{SODA
  '03})}. \bibinfo{publisher}{Society for Industrial and Applied Mathematics},
  \bibinfo{address}{USA}, \bibinfo{pages}{599–608}.
\newblock
\showISBNx{0898715385}


\bibitem[\protect\citeauthoryear{Erickson}{Erickson}{2012}]%
        {erickson2012combinatorial}
\bibfield{author}{\bibinfo{person}{Jeff Erickson}.}
  \bibinfo{year}{2012}\natexlab{}.
\newblock \showarticletitle{Combinatorial optimization of cycles and bases}.
\newblock \bibinfo{journal}{\emph{Advances in Applied and Computational
  Topology}}  \bibinfo{volume}{70} (\bibinfo{year}{2012}),
  \bibinfo{pages}{195--228}.
\newblock


\bibitem[\protect\citeauthoryear{Erickson and Whittlesey}{Erickson and
  Whittlesey}{2005}]%
        {Erickson2005}
\bibfield{author}{\bibinfo{person}{Jeff Erickson} {and} \bibinfo{person}{Kim
  Whittlesey}.} \bibinfo{year}{2005}\natexlab{}.
\newblock \showarticletitle{Greedy Optimal Homotopy and Homology Generators}.
  In \bibinfo{booktitle}{\emph{Proceedings of the Sixteenth Annual ACM-SIAM
  Symposium on Discrete Algorithms}} \emph{(\bibinfo{series}{SODA '05})}.
  \bibinfo{publisher}{Society for Industrial and Applied Mathematics},
  \bibinfo{address}{USA}, \bibinfo{pages}{1038–1046}.
\newblock
\showISBNx{0898715857}


\bibitem[\protect\citeauthoryear{Guskov and Wood}{Guskov and Wood}{2001}]%
        {Guskov2001}
\bibfield{author}{\bibinfo{person}{Igor Guskov} {and}
  \bibinfo{person}{Zo\"{e}~J. Wood}.} \bibinfo{year}{2001}\natexlab{}.
\newblock \showarticletitle{Topological Noise Removal}. In
  \bibinfo{booktitle}{\emph{Proceedings of Graphics Interface 2001}}
  \emph{(\bibinfo{series}{GI '01})}. \bibinfo{publisher}{Canadian Information
  Processing Society}, \bibinfo{address}{CAN}, \bibinfo{pages}{19–26}.
\newblock
\showISBNx{0968880800}


\bibitem[\protect\citeauthoryear{Hilaga, Shinagawa, Komura, and Kunii}{Hilaga
  et~al\mbox{.}}{2001}]%
        {Hilaga}
\bibfield{author}{\bibinfo{person}{Masaki Hilaga}, \bibinfo{person}{Yoshihisa
  Shinagawa}, \bibinfo{person}{Taku Komura}, {and} \bibinfo{person}{Tosiyasu
  Kunii}.} \bibinfo{year}{2001}\natexlab{}.
\newblock \showarticletitle{Topology matching for fully automatic similarity
  estimation of 3D shapes}.
\newblock \bibinfo{journal}{\emph{ACM SIGGRAPH}}, \bibinfo{pages}{203--212}.
\newblock
\urldef\tempurl%
\url{https://doi.org/10.1145/383259.383282}
\showDOI{\tempurl}


\bibitem[\protect\citeauthoryear{Kutz}{Kutz}{2006}]%
        {kutz2006}
\bibfield{author}{\bibinfo{person}{Martin Kutz}.}
  \bibinfo{year}{2006}\natexlab{}.
\newblock \showarticletitle{Computing Shortest Non-Trivial Cycles on Orientable
  Surfaces of Bounded Genus in Almost Linear Time}. In
  \bibinfo{booktitle}{\emph{Proceedings of the Twenty-Second Annual Symposium
  on Computational Geometry}} \emph{(\bibinfo{series}{SCG '06})}.
  \bibinfo{publisher}{Association for Computing Machinery},
  \bibinfo{address}{New York, NY, USA}, \bibinfo{pages}{430–438}.
\newblock
\showISBNx{1595933409}
\urldef\tempurl%
\url{https://doi.org/10.1145/1137856.1137919}
\showDOI{\tempurl}


\bibitem[\protect\citeauthoryear{Munkres}{Munkres}{1984}]%
        {munkres1984elements}
\bibfield{author}{\bibinfo{person}{J.R. Munkres}.}
  \bibinfo{year}{1984}\natexlab{}.
\newblock \bibinfo{booktitle}{\emph{Elements Of Algebraic Topology}}.
\newblock \bibinfo{publisher}{Perseus Publishing}.
\newblock
\showISBNx{9780201054873}
\showLCCN{lc84006250}


\bibitem[\protect\citeauthoryear{Pascucci, Scorzelli, Bremer, and
  Mascarenhas}{Pascucci et~al\mbox{.}}{2007}]%
        {Pascucci2007}
\bibfield{author}{\bibinfo{person}{Valerio Pascucci}, \bibinfo{person}{Giorgio
  Scorzelli}, \bibinfo{person}{Peer-Timo Bremer}, {and} \bibinfo{person}{Ajith
  Mascarenhas}.} \bibinfo{year}{2007}\natexlab{}.
\newblock \showarticletitle{Robust On-line Computation of Reeb Graphs:
  Simplicity and Speed}. In \bibinfo{booktitle}{\emph{ACM SIGGRAPH 2007
  Papers}} \emph{(\bibinfo{series}{SIGGRAPH '07})}. \bibinfo{publisher}{ACM},
  \bibinfo{address}{New York, NY, USA}, Article \bibinfo{articleno}{58}.
\newblock
\urldef\tempurl%
\url{https://doi.org/10.1145/1275808.1276449}
\showDOI{\tempurl}


\bibitem[\protect\citeauthoryear{{Shattuck} and {Leahy}}{{Shattuck} and
  {Leahy}}{2001}]%
        {Shattuck2001}
\bibfield{author}{\bibinfo{person}{D.~W. {Shattuck}} {and}
  \bibinfo{person}{R.~M. {Leahy}}.} \bibinfo{year}{2001}\natexlab{}.
\newblock \showarticletitle{Automated graph-based analysis and correction of
  cortical volume topology}.
\newblock \bibinfo{journal}{\emph{IEEE Transactions on Medical Imaging}}
  \bibinfo{volume}{20}, \bibinfo{number}{11} (\bibinfo{year}{2001}),
  \bibinfo{pages}{1167--1177}.
\newblock
\urldef\tempurl%
\url{https://doi.org/10.1109/42.963819}
\showDOI{\tempurl}


\bibitem[\protect\citeauthoryear{{Shinagawa}, {Kunii}, and
  {Kergosien}}{{Shinagawa} et~al\mbox{.}}{1991}]%
        {ReebAlgoIntroduction}
\bibfield{author}{\bibinfo{person}{Y. {Shinagawa}}, \bibinfo{person}{T.~L.
  {Kunii}}, {and} \bibinfo{person}{Y.~L. {Kergosien}}.}
  \bibinfo{year}{1991}\natexlab{}.
\newblock \showarticletitle{Surface coding based on Morse theory}.
\newblock \bibinfo{journal}{\emph{IEEE Computer Graphics and Applications}}
  \bibinfo{volume}{11}, \bibinfo{number}{5} (\bibinfo{date}{Sep.}
  \bibinfo{year}{1991}), \bibinfo{pages}{66--78}.
\newblock
\showISSN{1558-1756}
\urldef\tempurl%
\url{https://doi.org/10.1109/38.90568}
\showDOI{\tempurl}


\bibitem[\protect\citeauthoryear{Soman, Kothapalli, and Narayanan}{Soman
  et~al\mbox{.}}{2010}]%
        {FastCCL}
\bibfield{author}{\bibinfo{person}{Jyothish Soman}, \bibinfo{person}{Kishore
  Kothapalli}, {and} \bibinfo{person}{P~J Narayanan}.}
  \bibinfo{year}{2010}\natexlab{}.
\newblock \showarticletitle{Some GPU Algorithms For Graph Connected Components
  and Spanning Tree}.
\newblock \bibinfo{journal}{\emph{Parallel Processing Letters}}
  \bibinfo{volume}{20}, \bibinfo{number}{04} (\bibinfo{year}{2010}),
  \bibinfo{pages}{325--339}.
\newblock
\urldef\tempurl%
\url{https://doi.org/10.1142/S0129626410000272}
\showDOI{\tempurl}
\showeprint{https://doi.org/10.1142/S0129626410000272}


\bibitem[\protect\citeauthoryear{Tierny, Favelier, Levine, Gueunet, and
  Michaux}{Tierny et~al\mbox{.}}{2017}]%
        {ttk}
\bibfield{author}{\bibinfo{person}{Julien Tierny}, \bibinfo{person}{Guillaume
  Favelier}, \bibinfo{person}{Joshua~A. Levine}, \bibinfo{person}{Charles
  Gueunet}, {and} \bibinfo{person}{Michael Michaux}.}
  \bibinfo{year}{2017}\natexlab{}.
\newblock \showarticletitle{The {T}opology {T}ool{K}it}.
\newblock \bibinfo{journal}{\emph{IEEE Transactions on Visualization and
  Computer Graphics (Proc. of IEEE VIS)}} (\bibinfo{year}{2017}).
\newblock


\bibitem[\protect\citeauthoryear{Voss and Gerstein}{Voss and Gerstein}{2010}]%
        {voss20103v}
\bibfield{author}{\bibinfo{person}{Neil~R Voss} {and} \bibinfo{person}{Mark
  Gerstein}.} \bibinfo{year}{2010}\natexlab{}.
\newblock \showarticletitle{3V: cavity, channel and cleft volume calculator and
  extractor}.
\newblock \bibinfo{journal}{\emph{Nucleic acids research}}
  \bibinfo{volume}{38}, \bibinfo{number}{suppl\_2} (\bibinfo{year}{2010}),
  \bibinfo{pages}{W555--W562}.
\newblock


\bibitem[\protect\citeauthoryear{Wood, Hoppe, Desbrun, and Schr\"{o}der}{Wood
  et~al\mbox{.}}{2004}]%
        {wood2004}
\bibfield{author}{\bibinfo{person}{Zo\"{e} Wood}, \bibinfo{person}{Hugues
  Hoppe}, \bibinfo{person}{Mathieu Desbrun}, {and} \bibinfo{person}{Peter
  Schr\"{o}der}.} \bibinfo{year}{2004}\natexlab{}.
\newblock \showarticletitle{Removing Excess Topology from Isosurfaces}.
\newblock \bibinfo{journal}{\emph{ACM Trans. Graph.}} \bibinfo{volume}{23},
  \bibinfo{number}{2} (\bibinfo{date}{April} \bibinfo{year}{2004}),
  \bibinfo{pages}{190–208}.
\newblock
\showISSN{0730-0301}
\urldef\tempurl%
\url{https://doi.org/10.1145/990002.990007}
\showDOI{\tempurl}


\bibitem[\protect\citeauthoryear{Zayer, Mlakar, Steinberger, and Seidel}{Zayer
  et~al\mbox{.}}{2018}]%
        {zayer2018}
\bibfield{author}{\bibinfo{person}{Rhaleb Zayer}, \bibinfo{person}{Daniel
  Mlakar}, \bibinfo{person}{Markus Steinberger}, {and}
  \bibinfo{person}{Hans-Peter Seidel}.} \bibinfo{year}{2018}\natexlab{}.
\newblock \showarticletitle{Layered Fields for Natural Tessellations on
  Surfaces}.
\newblock \bibinfo{journal}{\emph{ACM Trans. Graph.}} \bibinfo{volume}{37},
  \bibinfo{number}{6}, Article \bibinfo{articleno}{264} (\bibinfo{date}{Dec.}
  \bibinfo{year}{2018}), \bibinfo{numpages}{15}~pages.
\newblock
\showISSN{0730-0301}
\urldef\tempurl%
\url{https://doi.org/10.1145/3272127.3275072}
\showDOI{\tempurl}


\bibitem[\protect\citeauthoryear{Zhou, Ju, and Hu}{Zhou et~al\mbox{.}}{2007}]%
        {Zhou2007}
\bibfield{author}{\bibinfo{person}{Qian-Yi Zhou}, \bibinfo{person}{Tao Ju},
  {and} \bibinfo{person}{Shi-Min Hu}.} \bibinfo{year}{2007}\natexlab{}.
\newblock \showarticletitle{Topology Repair of Solid Models Using Skeletons}.
\newblock \bibinfo{journal}{\emph{IEEE Transactions on Visualization and
  Computer Graphics}} \bibinfo{volume}{13}, \bibinfo{number}{4}
  (\bibinfo{date}{July} \bibinfo{year}{2007}), \bibinfo{pages}{675–685}.
\newblock
\showISSN{1077-2626}
\urldef\tempurl%
\url{https://doi.org/10.1109/TVCG.2007.1015}
\showDOI{\tempurl}


\end{thebibliography}

\end{document}